\documentclass[aps, prl,reprint,amsmath,amssymb]{revtex4-1}
\usepackage[ansinew]{inputenc}
\usepackage{graphicx}
\usepackage{textcomp}
\usepackage{pifont}
\usepackage{hyperref}
\usepackage[all]{hypcap}
\input{glyphtounicode}
\newcommand{\bra}[1]{\langle #1 |}
\newcommand{\ket}[1]{|#1\rangle}
\newcommand{\C}{^{13} \text{C}}

\begin{document}

\title{Robust quantum-network memory using decoherence-protected subspaces of nuclear spins}

\author{Andreas Reiserer}
\email{a.a.reiserer@tudelft.nl}
\thanks{A.R. and N.K. contributed equally to this work}
\affiliation{Kavli Institute of Nanoscience Delft, Delft University of Technology, PO Box 5046, 2600 GA Delft, The Netherlands}
\affiliation{QuTech, Delft University of Technology, PO Box 5046, 2600 GA Delft, The Netherlands}
\author{Norbert Kalb}
\email{a.a.reiserer@tudelft.nl}
\thanks{A.R. and N.K. contributed equally to this work}
\affiliation{Kavli Institute of Nanoscience Delft, Delft University of Technology, PO Box 5046, 2600 GA Delft, The Netherlands}
\affiliation{QuTech, Delft University of Technology, PO Box 5046, 2600 GA Delft, The Netherlands}
\author{Machiel S. Blok}
\affiliation{Kavli Institute of Nanoscience Delft, Delft University of Technology, PO Box 5046, 2600 GA Delft, The Netherlands}
\affiliation{QuTech, Delft University of Technology, PO Box 5046, 2600 GA Delft, The Netherlands}
\author{Koen J. M. van Bemmelen}
\affiliation{Kavli Institute of Nanoscience Delft, Delft University of Technology, PO Box 5046, 2600 GA Delft, The Netherlands}
\affiliation{QuTech, Delft University of Technology, PO Box 5046, 2600 GA Delft, The Netherlands}
\author{Daniel J. Twitchen}
\affiliation{Element Six Innovation, Fermi Avenue, Harwell Oxford, Didcot, Oxfordshire OX11 0QR, United Kingdom}
\author{Matthew Markham}
\affiliation{Element Six Innovation, Fermi Avenue, Harwell Oxford, Didcot, Oxfordshire OX11 0QR, United Kingdom}
\author{Tim H. Taminiau}
\affiliation{Kavli Institute of Nanoscience Delft, Delft University of Technology, PO Box 5046, 2600 GA Delft, The Netherlands}
\affiliation{QuTech, Delft University of Technology, PO Box 5046, 2600 GA Delft, The Netherlands}
\author{Ronald Hanson}
\affiliation{Kavli Institute of Nanoscience Delft, Delft University of Technology, PO Box 5046, 2600 GA Delft, The Netherlands}
\affiliation{QuTech, Delft University of Technology, PO Box 5046, 2600 GA Delft, The Netherlands}

\begin{abstract}
The realization of a network of quantum registers is an outstanding challenge in quantum science and technology. We experimentally investigate a network node that consists of a single nitrogen-vacancy (NV) center electronic spin hyperfine-coupled to nearby nuclear spins. We demonstrate individual control and readout of five nuclear spin qubits within one node. We then characterize the storage of quantum superpositions in individual nuclear spins under repeated application of a probabilistic optical inter-node entangling protocol. We find that the storage fidelity is limited by dephasing during the electronic spin reset after failed attempts. By encoding quantum states into a decoherence-protected subspace of two nuclear spins we show that quantum coherence can be maintained for over 1000 repetitions of the remote entangling protocol. These results and insights pave the way towards remote entanglement purification and the realisation of a quantum repeater using NV center quantum network nodes.
\end{abstract}

\maketitle

\section{Introduction}

Linking multi-qubit nodes into a large-scale quantum network \cite{kimble_quantum_2008, duan_colloquium:_2010, sangouard_quantum_2011, reiserer_cavity-based_2015} will open up exciting opportunities ranging from fundamental tests \cite{bancal_quantum_2012} and enhanced time-keeping \cite{komar_quantum_2014} to applications in quantum computing and cryptography \cite{kimble_quantum_2008, nickerson_topological_2013, barz_demonstration_2012, ekert_ultimate_2014}. Pioneering experiments with atomic ensembles \cite{sangouard_quantum_2011}, single atoms trapped in vacuum \cite{duan_colloquium:_2010, hofmann_heralded_2012, reiserer_cavity-based_2015, hucul_modular_2015} and spins in solids \cite{gao_coherent_2015, hensen_loophole-free_2015, delteil_generation_2015} have demonstrated entanglement between two optically connected nodes. Directly extending these schemes to quantum networks involving many nodes and spanning large distances is hindered by unavoidable imperfections, including photon loss and local control errors, which cause the success probability and entanglement fidelity to decay rapidly both with number of nodes and with distance.

\begin{figure}
\includegraphics[width=\columnwidth]{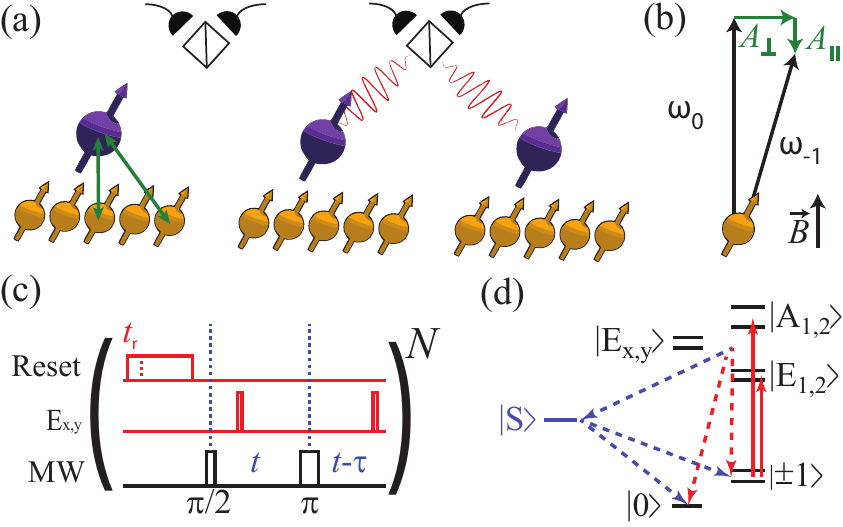}
\caption{\label{fig:setup}
Color) (a) Layered quantum network architecture. Individual electronic spins (purple spin symbols) are entangled probabilistically over large distances using photons (red curly lines). Each electronic spin is hyperfine coupled to a quantum register of surrounding nuclear spins (yellow) that can be deterministically controlled (green arrows). (b) Electron-nuclear coupling. The nuclear spins precess in an external magnetic field $\vec{B}$. The precession axis and frequency, $\omega_0$ or $\omega_{-1}$, (black vectors) depend on the state of the electronic spin, $\ket{0}$ or $\ket{-1}$, via the hyperfine interaction with parallel component $A_\parallel$ and perpendicular component $A_\perp$ (green vectors). (c) Experimental sequence to generate entanglement between remote NV electronic spins \cite{barrett_efficient_2005}. By optical pumping on the ``Reset" transition, the spin is initialized in $\ket{0}$ at time $t_r$. Subsequently, a spin superposition state is created via a microwave $\pi/2$ pulse. Spin-photon entanglement is then generated via two optical excitations, separated by a microwave $\pi$-pulse that inverts the spin state $\ket{0} \leftrightarrow \ket{-1}$. (d) NV electronic orbital and spin level scheme at cryogenic temperature. The ground states $\ket{0}$ ($\ket{\pm 1}$) are optically coupled to the excited states $\ket{E_{x,y}}$ ($\ket{E_{1,2}}$ and $\ket{A_{1,2}}$, red arrows), respectively. These states decay either directly (red dashed arrows) or via the metastable spin singlet states $\ket{S}$ (blue dashed arrows) to one of the ground states.
}
\end{figure}

These challenges can be overcome via entanglement purification \cite{bennett_purification_1996} in a repeater-type \cite{briegel_quantum_1998} setting that exploits quantum memories within each node \cite{childress_fault-tolerant_2006, fowler_surface_2010, nickerson_topological_2013}. Crucially, one needs to control and readout individual qubits within the node as well as create entanglement with remote qubits without inducing decoherence on the other qubits in the node. In principle, each of these tasks can be accomplished probabilistically using detectors and quantum memories for single optical photons \cite{duan_long-distance_2001}, but the resulting inefficiency poses a severe challenge for practical quantum network realizations \cite{duan_colloquium:_2010, sangouard_quantum_2011}. Instead, many efforts are geared towards a layered architecture, as depicted in Fig. \ref{fig:setup}(a). Here, remote quantum nodes are \emph{probabilistically} coupled via optical photons, while each node has several \emph{deterministically} addressable memory qubits that do not interact with and are therefore not disturbed by the optical channel.

A promising candidate for implementing such a quantum network architecture is the nitrogen-vacancy (NV) center in diamond. The NV electronic spin provides an optical interface that can be used to establish entanglement between distant nodes \cite{bernien_heralded_2013, pfaff_unconditional_2014, hensen_loophole-free_2015}, while nearby nuclear spins can serve as multi-qubit registers \cite{dutt_quantum_2007, robledo_high-fidelity_2011, liu_noise-resilient_2013, waldherr_quantum_2014} with second-long coherence times demonstrated in isolation \cite{maurer_room-temperature_2012}. Here, we address the open challenge of robust simultaneous execution of these two processes --- remote entanglement generation and local qubit storage ---  which is a key prerequisite for entanglement purification and quantum repeaters and therefore a critical task in quantum networks. We implement individual control over five nuclear spin qubits, in which we store quantum states while repeatedly using the electronic spin in a sequence which has previously been used to generate inter-node entanglement \cite{bernien_heralded_2013, pfaff_unconditional_2014, hensen_loophole-free_2015}.  We study in detail how the fidelity of storage depends on the coupling between electronic and nuclear spins and on the average time the electronic spin is in an unknown quantum state. We then use decoherence-protected subspaces to enhance the robustness of quantum state storage, which enables us to increase the exponential decay constant of the qubit fidelity above 1000 repetitions of the inter-node entangling sequence.

\
\section{Experimental setting}

The experiments are performed on a diamond device with a natural abundance of $\C$ nuclear spins ($I=1/2$, $m_I=1/2\equiv \ket{\uparrow}$, $m_I=-1/2\equiv \ket{\downarrow}$). We cool the device to a temperature of 4 K in a Helium bath cryostat and apply a magnetic field of 40 mT along the NV symmetry axis. Before each experimental run we prepare the NV center in the negative charge state and ensure resonance with the lasers \cite{robledo_high-fidelity_2011}. By using spin-selective optical transitions the electronic spin ($S=1$, $m_s=0\equiv \ket{0}$, $m_s=\pm 1 \equiv \ket{\pm 1}$) is initialized with a fidelity above 0.99 and read out in single shot with an average fidelity of about 0.94. We employ tailored pulse sequences on the electronic spin \cite{taminiau_universal_2014, cramer_repeated_2015} that yield high-fidelity individual control of five $\C$ spin qubits surrounding the NV center studied here (the same as in Ref. \cite{cramer_repeated_2015}). In Table I we list for each nuclear spin qubit the measured hyperfine coupling parameters, the dephasing time $T_2^*$ and the combined fidelity of spin initialization and readout $F_{i,r}$ (see S.I.).

\begin{table} [h!]
\label{tab:Carbons}
\begin{tabular}{c || c | c | c | c | c }
$\C$ number & $\frac{\Delta \omega}{2 \pi}$ (kHz) & $ A_\parallel$ (kHz)& $ A_\perp$ (kHz) & $T_2^*$ (ms) & $F_{i,r}$ \\
\hline
1           & -15.4                & -11.0          & 55           & 6(1)         & 0.89(2) \\ 
2           & 18.4                 &  21.2          & 43           & 13(1)         & 0.96(2) \\ 
3           & 23.7                 &  24.7          & 26           & 19(2)         & 0.97(2) \\ 
4           & -37.0                & -36.0          & 25           & 10(1)         & 0.92(2) \\ 
5           & -48.6                & -48.7          & 12           & 4(1)         & 0.90(2) \\ 
\end{tabular}
\caption{Experimentally determined parameters of the five $\C$ nuclear spin qubits used in this work: hyperfine couplings $A_\parallel$ and $A_\perp$, precession frequency difference $\Delta \omega$, and combined initialization and readout fidelity $F_{i,r}$. The hyperfine parameters are measured via dynamical decoupling spectroscopy \cite{taminiau_detection_2012} and have an uncertainty in the last digit.
}
\end{table}

We now focus on nuclear spin coherence during application of the Barrett-Kok inter-node entangling sequence \cite{barrett_efficient_2005} (see Fig. 1c) that was used in recent experiments \cite{bernien_heralded_2013, pfaff_unconditional_2014, hensen_loophole-free_2015}. This protocol has two steps. First, entanglement between the electronic spin and the emission time of a single photon is created at both nodes. Subsequently the two photons are measured after interfering on a beamsplitter, probabilistically projecting the electronic spins into a maximally entangled state. Because of unavoidable inefficiencies including photon loss, this sequence has to be repeated many times to generate remote entanglement with a high probability.

Each time an entanglement attempt fails the electronic spin of the NV center is projected into an unknown state and has to be reset. This can lead to decoherence of the surrounding nuclear spin quantum memories via the (always-on) hyperfine interaction. The interaction Hamiltonian is in secular approximation:
\begin{equation}
H_{hf}/2\pi= A_{\parallel} S_z I_z + A_{\perp} S_z I_x.
\end{equation}
Here, $S$ and $I$ denote the electronic and nuclear spin operators, respectively, and $A_\parallel$ ($A_\perp$) denote the parallel (perpendicular) hyperfine coupling strength. If the electronic spin state is $\ket{0}$, the nuclear spin will precess at the Larmor frequency $\omega_0 = 2 \pi \times \gamma |\vec{B}|$ around the applied magnetic field $\vec{B}$ (see Fig. \ref{fig:setup}(b)), where $\gamma=11$ kHz/mT is the nuclear gyromagnetic ratio. If the electronic spin state is $\ket{-1}$, however, the nuclear spin will precess around a slightly tilted axis and at a different frequency $\omega_{-1} = 2 \pi \times \sqrt{(\gamma|\vec{B}|+A_\parallel)^2+A_\perp^2}$. In a sufficiently strong magnetic field, $\gamma |\vec{B}| \gg \sqrt{A_\perp^2+A_\parallel^2} $, the change in precession axis is quadratically suppressed and nuclear spin decoherence is mainly caused by dephasing due to the linear change in the precession frequency
\begin{equation}
\Delta \omega = \omega_{0} - \omega_{-1} \simeq 2 \pi \times A_\parallel.
\end{equation}
Thus, randomisation of the electronic spin state is expected to lead to dephasing of a nuclear spin on a timescale that is inversely proportional to the parallel hyperfine coupling strength \cite{jiang_coherence_2008, blok_towards_2015}.

This dephasing can be mitigated by a suited dynamical decoupling sequence \cite{blok_towards_2015}, which is inherent in the Barrett-Kok entangling sequence (Fig. \ref{fig:setup}(c)): for $\tau=0$, the time interval between the microwave (MW) $\pi/2$ pulse and the MW $\pi$ pulse has the same duration as the time interval between the MW $\pi$ pulse and the start of the electronic spin reset. Thus, the electronic spin will be in $\ket{0}$ and $\ket{-1}$ for an equal amount of time, irrespective of the random optical projection. Thus, under the condition that the spin reset is instantaneous and happens at a precisely known time, the dephasing is exactly nullified. However, electronic spin reset by optical pumping is a stochastic process of finite time duration. As the spin state is not known during this process, it induces irreversible dephasing of the nuclear spins. Therefore, it is desirable to use nuclear spins with weak coupling strength and to make the electronic spin reset as short as possible.

\section{Electronic spin reset}

We first investigate the spin reset process and optimize the reset time. We initialize the electronic spin in $\ket{0}$ by applying laser fields that are on resonance either with the transitions $\ket{-1} \leftrightarrow \ket{E_1}$ and $\ket{+1} \leftrightarrow \ket{E_2}$ or, for comparison, with the transitions $\ket{-1} \leftrightarrow \ket{A_1}$ and $\ket{+1} \leftrightarrow \ket{A_2}$, see Fig. \ref{fig:setup}(d). Compared to our previous remote entanglement experiments \cite{bernien_heralded_2013, pfaff_unconditional_2014, hensen_loophole-free_2015}, the use of higher magnetic fields requires a second laser beam because of the comparably large ground state Zeeman splitting between $\ket{-1}$ and $\ket{+1}$ of about $2\,\text{GHz}$. The lasers address different excited states to avoid pumping to a coherent dark state. Both repump laser beams have approximately circular polarization and equal power. The excited states have a lifetime on the order of $10\,\text{ns}$ \cite{goldman_phonon-induced_2015}. From the excited states, the NV can decay either back to $\ket{\pm1}$, or to the metastable spin singlet states $\ket{S}$. In addition, spin mixing in the excited states also opens a direct decay channel to $\ket{0}$ \cite{doherty_nitrogen-vacancy_2013}.

\begin{figure}
\includegraphics[width=\columnwidth]{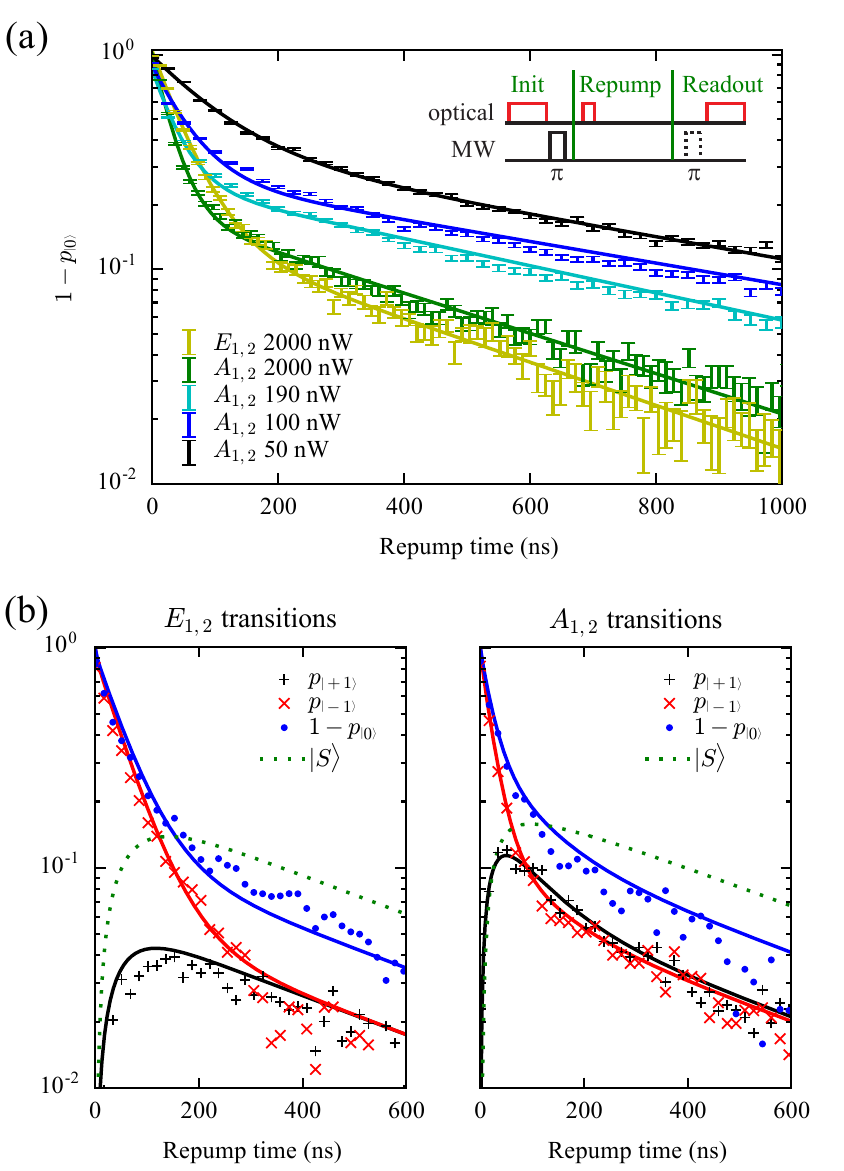}
\caption{\label{fig:optical_pumping}
Color) NV electronic spin initialization. (a) Probability that the electronic spin is pumped to $\ket{0}$ as a function of the repump laser pulse duration when the spin is initially prepared in $\ket{-1}$. The inset shows the used pulse sequence. We observe a double exponential decay (solid fit curves), with a timescale and relative amplitude that depends on the used ``Reset" transition. Reduced laser power leads to slower initialization timescales. (b) Probability that the NV is found in state $\ket{0}$, $\ket{+1}$, and $\ket{-1}$ for 2000 nW repump power in the $\ket{E_{1,2}}$ (left) or $\ket{A_{1,2}}$ (right) configuration. The solid lines are calculated using a rate equation model described in the text. For long repumping time, the calculated population in the metastable singlet states (green dashed line) dominates the repumping process.}
\end{figure}

To determine the time it takes to reset the spin, we prepare it in $\ket{-1}$ and pump it with the reset lasers for a varying duration. After a delay of $2.5\,\mu\text{s}$ to ensure that no population is left in the excited or singlet states \cite{doherty_nitrogen-vacancy_2013}, we measure the population in $\ket{0}$, see Fig. \ref{fig:optical_pumping}(a). The spin pumping process exhibits a double-exponential decay with a relative amplitude ratio for the fast and slow time scales of around 3:1, which slightly depends on the excited states used. The two timescales of this decay depend on the applied laser power. At high power, the reset timescales saturate (see S.I.) at $29(1)\,\text{ns}$ and $463(14)\,\text{ns}$ when driving transitions to $\ket{A_{1,2}}$ (green) and at $48(1)\,\text{ns}$ and $432(26)\,\text{ns}$ when driving transitions to $\ket{E_{1,2}}$ (yellow).

We attribute the slower timescale, which is the same for both configurations within measurement uncertainty, to the decay constant of the metastable singlet states. The fitted value is in the same range as previously reported values \cite{robledo_control_2010}. The faster timescale has a contribution from both direct spin flip transitions to $\ket{0}$ and transitions into the singlet states which then decay to $\ket{0}$. The difference in fast timescales between the two configurations is explained by different decay rates to the ground states and metastable singlet states from the excited states used \cite{goldman_phonon-induced_2015}. When the laser power is reduced, we observe a gradual increase of both timescales, as shown for the $\ket{A_{1,2}}$ configuration (cyan, blue and black curves).

To obtain additional insight into the spin reset process, we measure the probabilities $p_i$ to arrive in the states $i=\ket{0}, \ket{-1}$, and $\ket{+1}$, again $2.5\,\mu\text{s}$ after applying a repump pulse of varying duration. In Fig. \ref{fig:optical_pumping}(b) we plot $1-p_{\ket{0}}$, $p_{\ket{-1}}$, and $p_{\ket{+1}}$. We fit the data to rate equation models (solid lines), one for each repump configuration. These models assume the lifetimes of the individual states as measured in Ref. \cite{goldman_phonon-induced_2015}, a lifetime of the metastable singlet states of 440 ns (taken from the above fits), a relative singlet decay ratio $R_i$ to the states $i$ of $(R_{\ket{0}}: R_{\ket{+1}}: R_{\ket{-1}})=(2:1:1)$ \cite{doherty_nitrogen-vacancy_2013}, and full decay of the singlet and excited states before the ground state population is measured. The model uses equal decay of both excited states $\ket{A_{1,2}}$ to $\ket{\pm1}$, while the states $\ket{E_{1,2}}$ decay either to $\ket{+1}$ or to $\ket{-1}$ \cite{doherty_nitrogen-vacancy_2013}. The decay rates of the excited states to $\ket{0}$ and the rate of excitation and stimulated emission caused by the repump laser are free parameters in the model. The quantitative agreement between data and the model strengthens the hypothesis that the slow timescale of the repump process is dominated by the lifetime of the metastable singlet states. The population of the metastable singlet states (before decay to the ground states) predicted by the models is shown as the green dashed curve.

\section{Single-nuclear-spin quantum memory}

We now turn to the dephasing of single nuclear spin qubits induced by the stochastic trajectory of the electronic spin state during reset. Using recently developed techniques \cite{cramer_repeated_2015}, we first initialize one of the five controlled nuclear spins by a projective measurement (see S.I.). Ideally, this prepares the nuclear spin superposition state $\frac{1}{\sqrt{2}}(\ket{\downarrow}+\ket{\uparrow})$. We then perform $N$ repetitions of the remote entanglement sequence. We omit the short optical $\pi$ pulses, as they are expected to induce negligible nuclear spin dephasing since they preserve the electronic spin eigenstate and can be timed such that the detrimental effect of occasional spin flips ($p_{Flip}<1\,\%$) \cite{robledo_high-fidelity_2011} is mitigated by the dynamical decoupling sequence. In addition, the fast optical pulses only lead once per sequence to a population of the excited state, whose spin projection is zero and whose $12\,\text{ns}$ lifetime is short compared to the reset procedure. We track the dephasing of the nuclear spins by measuring the length of their Bloch vector projection onto the XY plane. We do not include the Z projection as it remains small. In addition, we discard the small fraction of experimental runs in which the NV electronic spin is ionized (S.I.).

\begin{figure}
\includegraphics[width=\columnwidth]{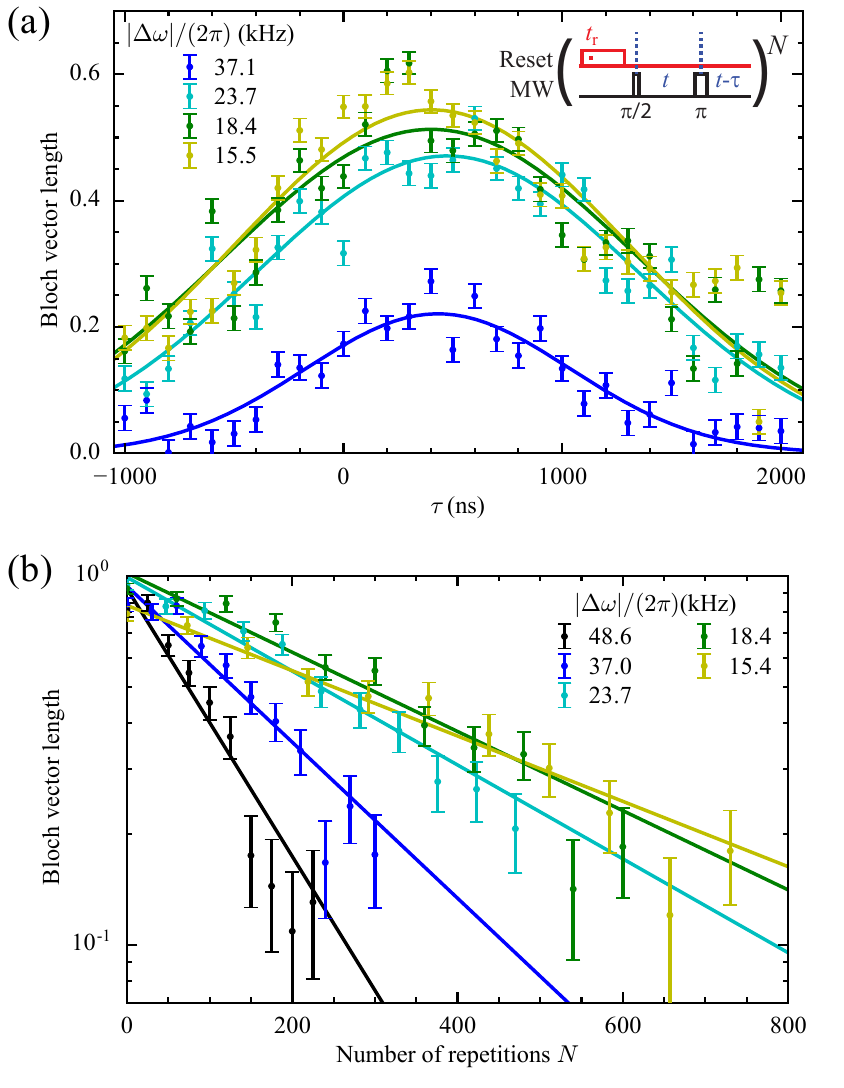}
\caption{\label{fig:fidelity_vs_rep}
Color) Dephasing of $\C$ nuclear spins. (a) Optimization of the dynamical decoupling sequence timing. Different nuclear spins (colors) are initialized in a balanced superposition state and the entanglement sequence is performed $N=200$ times. The duration of the second wait interval is swept and the length of the Bloch vector XY projection is measured. All measured nuclear spins exhibit the same broad optimum around $0.4\,\mu$s, as can be seen from the Gaussian fit curves. (b) Dephasing of nuclear spins when the number of random reset processes is increased. The data shows measurement results of all five individually controlled nuclear spins, where increasing coupling strength leads to steeper decay curves. The solid lines are exponential fits.
}
\end{figure}

We first investigate and optimize the timings $t$ and $\tau$ of the dynamical decoupling sequence shown in Fig. \ref{fig:setup}(c). We find that the dephasing rate shows a clear minimum when $t$ equals the Larmor period of the nuclear spins (see S.I.), in which case the effect of entanglement between electronic and nuclear spins caused by the tilted rotation axis, as shown in Fig. \ref{fig:setup}(b), is minimized. We therefore set $t = \frac{2 \pi}{\omega_{0}} \simeq 2.3\,\mu\text{s}$. We then repeat the entanglement sequence $N=200$ times and vary the time $\tau$. Assuming an exponential distribution of the repumping time $t_r$, one expects to obtain the best possible decoupling when $\tau$ is equal to $\langle t_r \rangle$ \cite{blok_towards_2015}. As can be seen in Fig. \ref{fig:fidelity_vs_rep}(a), we observe a broad Gaussian distribution centered around an optimal value of $\tau \approx 0.44\,\mu\text{s}$ for all four measured nuclear spins, in good agreement with the slow timescale of the initialization process shown in Fig. \ref{fig:optical_pumping}(a). As mentioned, we attribute this timescale to the decay of the metastable singlet states. At first sight, it is surprising that a singlet state which has zero spin projection and thus no hyperfine coupling would induce dephasing. A possible explanation is that the orbital angular momentum of an E-symmetric singlet state induces a magnetic moment that is comparable to that of the electronic spin ground states.

Additional dephasing can result from experimental imperfections. To prevent errors caused by imperfect spin initialization, e.g. when the laser power drifts over time, we apply the repump laser longer than required for the initialization curves to saturate below 0.01, which is $2 \,\mu$s ($1.5 \,\mu$s) for the $\ket{E_{1,2}}$($\ket{A_{1,2}}$) repump configuration at 2000 nW, respectively. To prevent errors caused by imperfect MW pulses, we employ a Hermite pulse envelope with a broad spectrum in order to drive the $\ket{0} \leftrightarrow \ket{-1}$ transition irrespective of the spin state of the nitrogen nucleus of the NV center. We employ this pulse in a BB1 composite pulse sequence \cite{vandersypen_nmr_2005}, consisting of five pulses of less than $0.2\,\mu\text{s}$ duration each, to further suppress pulse errors. In this way, we are able to exclude experimental imperfections as a relevant source of the observed dephasing (see S.I. for details).

With the optimized timings and pulses, we then vary the number of repetitions $N$ of the remote entanglement sequence. For all five nuclear spins, we observe an exponential decay of the single-qubit coherence with $N$, see Fig. \ref{fig:fidelity_vs_rep}(b). Even for the nuclear spin with the largest parallel hyperfine coupling --- for which the dephasing is fastest ---  more than $N=100$ repetitions of the entanglement sequence can be run before the Bloch vector length drops to 1/e. For the nuclear spin with the smallest coupling strength, this number increases to $N \approx 500$. Further improvements could be achieved by using nuclear spins with even lower parallel hyperfine coupling, although this would generally come at the price of an increased duration of local control operations.

\section{Decoherence-protected subspace quantum memory}

Motivated by the observation that the memory dephasing is mainly determined by the parallel hyperfine coupling strength, we investigate a different approach to increasing the maximum number of repetitions before a qubit is dephased. Instead of encoding the qubit in a single nuclear spin, we can encode in a decoherence-protected subspace (DPS) \cite{lidar_decoherence-free_1998} of two or more nuclear spins. In this way, the net parallel hyperfine coupling can be strongly reduced while the speed of the individual quantum gates remains the same.

A natural choice for a DPS with reduced dephasing is given by the basis states $\ket{\downarrow_i \uparrow_j}$ and $\ket{\uparrow_i \downarrow_j}$ of nuclear spins $i$ and $j$ with comparable parallel hyperfine coupling strength. An encoded qubit will then to first order dephase at a rate that is determined by the coupling strength \emph{difference} $\Delta \omega \simeq 2 \pi \times (A_{\parallel,i}-A_{\parallel,j})$, which can be much smaller than the individual coupling strengths. On the other hand, encoding a qubit in the states $\ket{\uparrow_i \uparrow_j}$ and $\ket{\downarrow_i \downarrow_j}$ will lead to increased dephasing rates.

To experimentally verify these expectations, we create the states $(\ket{\downarrow_i \uparrow_j} + \ket{\uparrow_i \downarrow_j})/\sqrt{2}$ and $(\ket{\uparrow_i \uparrow_j} + \ket{\downarrow_i \downarrow_j})/\sqrt{2}$ \cite{cramer_repeated_2015} (see S.I.) and measure the qubit state projection onto the XY plane of the Bloch sphere under the remote entangling protocol. Fig. \ref{fig:reps_vs_coup}(a) shows the results obtained for nuclear spins 2 and 3. When initializing the qubit in a decoherence-protected (decoherence-enhancing) two-spin state, we observe a strong improvement (reduction) of the maximum number of repetitions. In the DPS case, we can perform more than one thousand repetitions before the Bloch vector length drops to 1/e. This decay constant can be fully explained by the intrinsic dephasing time $T_2^*$ of the nuclear spins (see S.I.). This shows that the dephasing induced by the entanglement protocol has become negligible in this DPS quantum memory.

The coherence of a DPS with small effective coupling strength might also be limited by the population decay ($T_1$) of the individual spins induced by the entanglement protocol. For the current sample, we find an exponential decay of population with $N$ with decay constants in the range of 1000 to 10000 repetitions, depending on the individual nuclear spins used (see S.I.). We note that this effect is not limiting the coherence of the two-qubit DPS investigated here, but we expect it to become relevant for the smallest investigated coupling strengths once the intrinsic dephasing ($T_2^*$) is canceled by an echo.

\section{Scaling of the dephasing rate with hyperfine coupling strength and repump duration}

\begin{figure}
\includegraphics[width=\columnwidth]{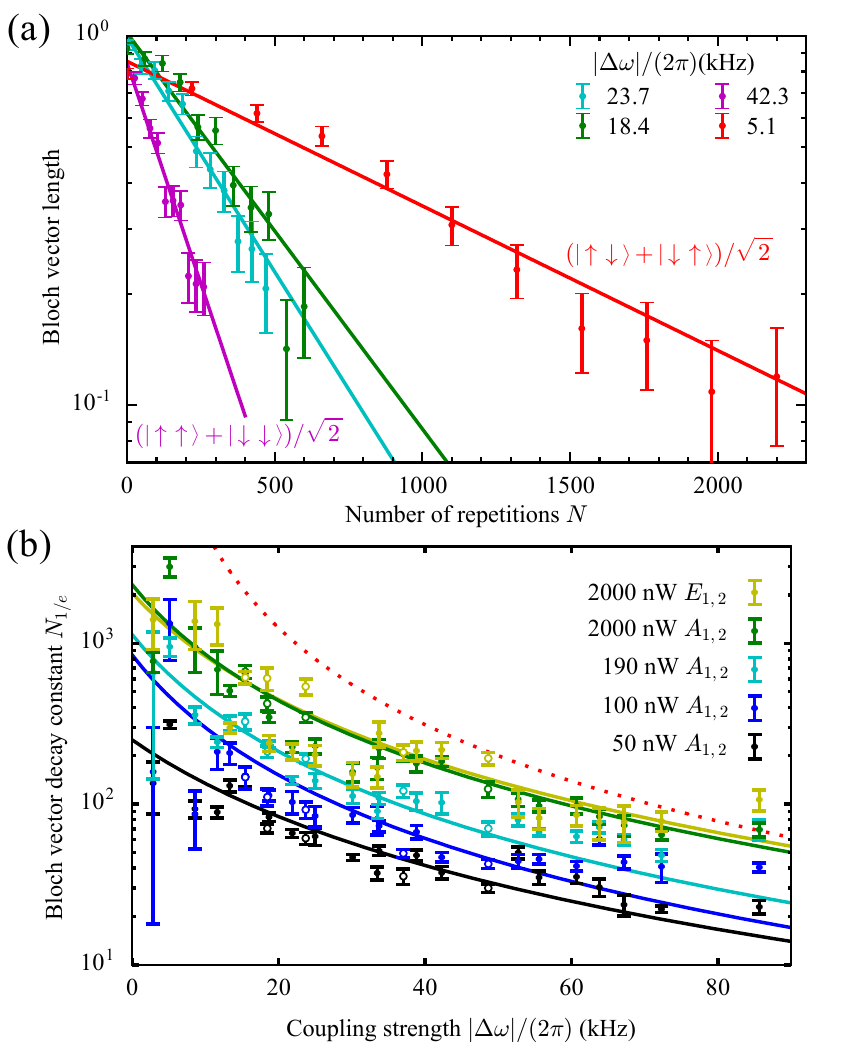}
\caption{\label{fig:reps_vs_coup}
Color). Encoding of a quantum bit in decoherence-protected subspaces. (a) Encoding in nuclear spin 2 or 3 (cyan and green) shows similar decay with increasing $N$. Encoding in a decoherence-protected (decoherence-enhanced) subspace leads to strongly decreased (enhanced) dephasing shown in red (magenta). The initial fidelity in the two-spin case is slightly reduced because encoding and readout require more control operations on the nuclear spins. (b) Number of sequence repetitions that are possible before the nuclear qubit Bloch vector length drops to 1/e of its initial value, for qubits encoded in single nuclear spins (empty circles) and in two-spin states (filled circles) of different effective coupling strengths $\Delta \omega$. The four depicted data sets are taken for increasing repump duration, caused by a reduced repump laser power. The solid curves are fits according to the model presented in the text.
}
\end{figure}

Finally, we perform an extensive quantitative investigation of the scaling of dephasing with the coupling strength and with the time it takes to reset the electronic spin. Fig. \ref{fig:reps_vs_coup}(b) shows the number of entanglement sequence repetitions $N_{1/e}$ for which the state fidelity of a balanced superposition state decays to 1/e of the initial value [i.e. the fitted decay constant in \ref{fig:reps_vs_coup}(a)]. We investigate all 5 individual nuclear spins (open circles) and all 20 possible two-spin subspaces (filled circles), whose coupling strength is the sum or difference of the individual ones. The depicted five data sets correspond to the different values of the repumping time constants shown in Fig. \ref{fig:optical_pumping}. To ensure that we only investigate the scaling of the dephasing with coupling strength and repump duration, we correct for the effects of $T_2^*$ decay, which becomes dominant for the leftmost three data points (see S.I.) and could be compensated by a suited echo sequence on the nuclear spins \cite{maurer_room-temperature_2012}.

We compare the data to the model of Blok et al. \cite{blok_towards_2015}, which assumes an exponentially distributed repump timescale and that the NV stays in the ground state $\ket{-1}$ until it is reset. When the value of $\tau$ used in the dynamical decoupling sequence is equal to the average repump time $\langle t_{r} \rangle$, a nuclear spin has acquired a phase shift of $\Delta \omega \times (t_{r} - \tau)$ until the electronic spin is reset to $\ket{0}$. In the limit of large $N$, the binomial probability distribution of required electronic spin resets can be approximated by a Gaussian distribution. For a balanced superposition input state, as investigated here, this leads to a predicted qubit fidelity of

\begin{equation}
\label{eq:fidelity_vs_rep}
F= \frac{1}{2} + \frac{1}{2^{N+1}} ( 1+\text{e}^{ - \Delta \omega^2 \tau^2 /2})^N.
\end{equation}

Thus, the model correctly predicts the observed exponential dephasing of the qubit with increasing number of repetitions $N$. As expected, the decay constant depends on the hyperfine coupling strength of the nuclear spin with a faster decay for increased coupling strength. However, the prediction of the model when inserting the measured optimal value of $\tau = 0.44\,\mu\text{s}$ (red dotted line) does not exhibit quantitative agreement with the measured data. A possible explanation is that the model neglects the double-exponential reset time distribution and the time spent in $\ket{+1}$, in one of the excited states or in the meta-stable singlet states. These assumptions are certainly not justified in the present experiment.

To account for this, we leave $\tau$ in Eq. \ref{eq:fidelity_vs_rep} as a free parameter and introduce an offset parameter $C$ to the coupling strength, $\Delta \omega \rightarrow (\Delta \omega + C)$. Setting $C$ to $\approx 2\pi \times 15$ kHz leads to reasonable agreement of the model (solid curves in Fig. \ref{fig:reps_vs_coup}(b)) with the data. The observed fit values ($\tau=0.43(3)\,\mu\text{s}$ for the $\ket{E_{1,2}}$ and $\tau=0.46(1)\,\mu\text{s}$ for the $\ket{A_{1,2}}$ repumping configuration) agree within error with both the measured slow repumping timescale,  see Fig. \ref{fig:optical_pumping}(a), and the optimal value of $\tau$ in Fig. \ref{fig:fidelity_vs_rep}(a).

\section{Conclusion and outlook}
We have studied a prototype quantum network node consisting of nuclear spin qubits hyperfine coupled to an optically active electronic spin in a diamond with natural isotope abundance. Repeated application of a remote entangling protocol is observed to cause dephasing of the nuclear spin qubits. We found that this dephasing can be mitigated by using nuclear spins with small parallel hyperfine coupling strengths, and, even more effectively, by encoding in decoherence-protected subspaces of multiple spins. For the smallest coupling strengths investigated here, the storage of quantum states is robust to more than 1000 remote entangling attempts. We expect that our experimental findings can be generalized to other physical systems \cite{awschalom_quantum_2013} and other quantum protocols in which a repeated reset of an ancilla qubit with always-on coupling to a memory is required.

In the future, the implementation of high quality optical cavities should allow for a further reduction of the time it takes to reset the electronic spin, as the Purcell effect \cite{purcell_spontaneous_1946} induced by such resonator increases the probability of direct spin-flips without populating the singlet states. In addition, the development of techniques to measure the electronic spin state non-destructively or within a decoupling sequence might fully eliminate the need for probabilistic repumping. Finally, the realization of quantum networking protocols that are based on photon absorption \cite{kalb_heralded_2015, yang_high_2015, jones_design_2015} rather than photon emission may reduce the number of required electronic spin resets until a successful entanglement event is heralded. Even in the absence of such future improvements, we anticipate that the current results will enable first demonstrations of the purification of remote entanglement \cite{bennett_purification_1996, campbell_measurement-based_2008} and proof-of-principle operation of a quantum repeater \cite{briegel_quantum_1998} based on NV centers in diamond \cite{childress_fault-tolerant_2006}.

\newpage
\section{Supplemental information}

\subsection{$\C$ spin initialization and readout}
\begin{figure}
\includegraphics[width=\columnwidth]{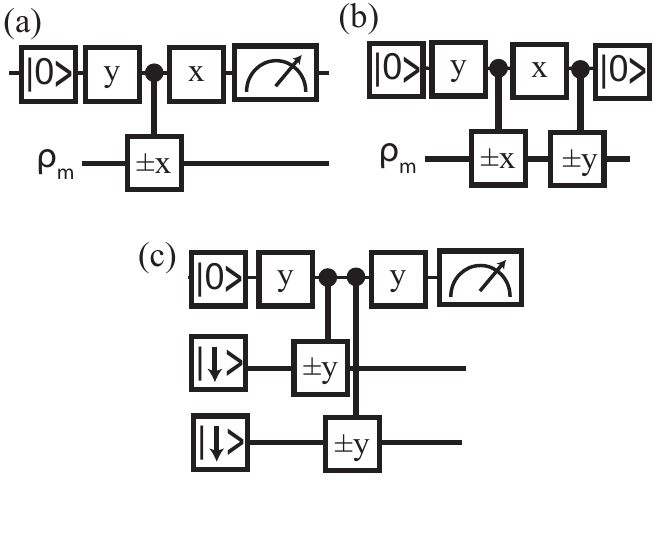}
\caption{ \label{fig:GateSequences}
Gate sequences used for initialization of $\C$ spin quantum memories. $x (y)$ denotes a MW pulse that rotates the electronic spin by an angle of $\pi/2$ around the $x (y)$ axis in the Bloch sphere picture. For an explanation of the used sequences, see text. (a) Initialization of a $\C$ nuclear spin in $(\ket{\uparrow}+\ket{\downarrow})/\sqrt{2}$. (b) Initialization in $\ket{\downarrow}$. (c) Initialization in  $(\ket{\uparrow\downarrow}+\ket{\downarrow\uparrow})/\sqrt{2}$ or $(\ket{\uparrow\uparrow}+\ket{\downarrow\downarrow})/\sqrt{2}$.
}
\end{figure}

In this section, we explain the procedures used to initialize and readout the $\C$ spins using the NV electronic spin as an ancilla \cite{cramer_repeated_2015}. We first describe how a single nuclear spin is prepared in the state $\ket{X} \equiv (\ket{\downarrow}+\ket{\uparrow})/\sqrt{2}$. We employ the gate sequence shown in Fig. \ref{fig:GateSequences}(a). The electronic spin is first reset to $\ket{0}$ using a laser pulse on the repump transitions (see main text). Subsequently, it is rotated to $(\ket{0}+\ket{1})/\sqrt{2}$ by a single $\pi/2$ microwave (MW) pulse. Then, a sequence of MW $\pi$ pulses performs a selective rotation on one of the nuclear spins, which is initially in a mixed state $\rho_m$. The rotation is conditional on the initial electronic spin state, as described in detail in Ref. \cite{taminiau_universal_2014}. In this way, the nuclear and electronic spins are correlated. After another MW $\pi/2$ pulse, a measurement of the electronic spin projects the nuclear spin to $\ket{\pm X}$, where the sign depends on the measurement result.

In our experiment, the readout fidelity of the ancilla qubit is asymmetric: If a photon is detected, the NV is with a very high probability ($>99\%$) in the fluorescent state $\ket{0}$. Therefore, we repeat the initialization sequence until $\ket{0}$ is measured, which prepares the nuclear spin in $\ket{X}$. The same procedure can be used to initialize all controlled nuclear spins sequentially.

The procedure to initialize a single nuclear spin in $\ket{\downarrow}$ is shown in Fig. \ref{fig:GateSequences}(b). Here, the electronic spin is prepared in $\ket{0}$. We then apply a reduced swap operation, which consists of two MW rotations of the electronic spin and two rotations of the nuclear spin that are conditional on the electronic spin. Ideally, this deterministically transfers the electronic spin $\ket{0}$ state to the nuclear spin state $\ket{\downarrow}$ \cite{cramer_repeated_2015}.

Finally, we describe the pulse sequence used to prepare a single qubit in a two-spin decoherence-protected subspace. Here, the nuclear spins are prepared in $\ket{\downarrow \downarrow}$ by sequential application of the scheme described above. Then, the gate sequence shown in Fig. \ref{fig:GateSequences}(c) prepares the nuclear spins either in $(\ket{\downarrow \downarrow}+\ket{\uparrow \uparrow})/\sqrt{2}$ or in $(\ket{\uparrow \downarrow}+\ket{\downarrow \uparrow})/\sqrt{2}$, again depending on the measurement result \cite{cramer_repeated_2015}. To avoid errors caused by the asymmetric electron readout fidelity, we again condition the start of the experiment on the detection of a photon, and we insert a MW $\pi$ pulse before readout if we want to prepare the other nuclear spin state.

The techniques to do tomography on the nuclear spin consist of the same set of operations as those for initialization. They are described in detail in the supplementary information of Ref. \cite{cramer_repeated_2015}.

To characterize the control we achieve over the five individual nuclear spins, we prepare each of them separately in $\ket{X}$ and then perform tomography using the electronic spin as an ancilla. We correct for ancilla readout errors. The combined nuclear spin initialization and readout fidelity $F_{i,r}$ is the overlap of the reconstructed nuclear spin density matrix $\rho$ with the ideally expected state, i.e. $F_{i,r}=\bra{X}\rho\ket{X}$. We emphasize that measuring $F_{i,r}$ involves several two-qubit quantum gates and single-qubit rotations. The individual fidelity of each of these operations is expected to be higher than the combined value.

\subsection{Optimization of the dynamical decoupling sequence}
As can be seen in Fig. 1(b) of the main text, the hyperfine coupling of the NV electronic and the $\C$ nuclear spins leads to two effects: first, a change of the nuclear spin precession frequency, and second, a tilt of the precession axis. In the limit of large magnetic fields, the latter is quadratically suppressed. At the magnetic field used in the experiments, however, the nuclear spins may become partly entangled with the electronic spin in each repetition of the entanglement sequence. This induces decoherence upon projection of the latter.

To minimize this decoherence, two strategies can be explored in order to minimize the entanglement. First, keeping the entanglement sequence very short by minimizing the time $t$ between the pulses, such that the overall nuclear spin evolution during the sequence remains small. Second, the timing of the sequence can be adjusted such that it is matched to one of the nuclear spin precession frequencies, i.e. $t = \frac{2 \pi}{\omega_{-1}}$ or $t = \frac{2 \pi}{\omega_{0}}$. In this case, the nuclear spins have approximately undergone two full rotations (one around the magnetic field and one around the tilted axis) and end up at the same projection along the $Z$ axis as they started, which minimizes entanglement between electronic and nuclear spin.

\begin{figure}
\includegraphics[width=\columnwidth]{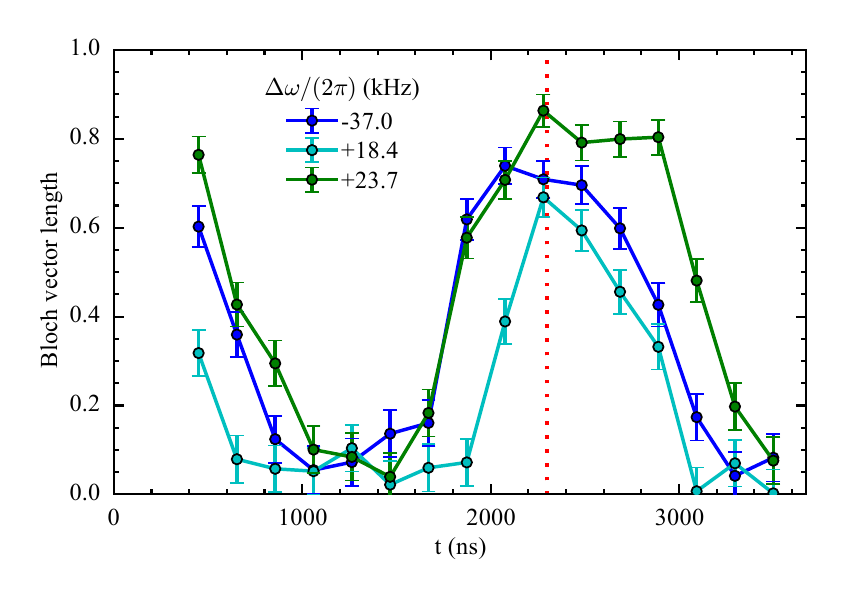}
\caption{ \label{fig:Optimizing_t}
Optimization of the dynamical decoupling sequence. Three single $\C$ nuclear spins are individually prepared along the $Z$ axis and the entanglement sequence is repeated $N=450$ times. The state of the encoded qubit is best preserved when the timing $t$ of the dynamical decoupling sequence matches the nuclear spin precession period (red dashed line).
}
\end{figure}

Fig. \ref{fig:Optimizing_t} shows a measurement in which this theoretical expectation is observed. The measurement uses the sequence shown in the inset of Fig. 3(a) in the main text. Single $\C$ spins are initialized in $\ket{\downarrow}$ and the entanglement sequence is applied $N= 450$ times. We then measure the length of the Bloch vector $Z$ projection for several values of the wait time $t$ in the dynamical decoupling sequence. We observe two distinct maxima, one at short $t$ and one around the inverse of the nuclear spin precession frequency. The position of the latter slightly depends on the sign and magnitude of the hyperfine coupling. We use $t=\frac{2 \pi}{\omega_{0}}\simeq 2.3\,\mu\text{s}$ (red dashed line) in all experiments presented in the main text.

\subsection{Influence of electronic spin reinitialization and MW pulse errors}
In addition to the sources of dephasing described in the main text, also experimental imperfections such as MW pulse errors and imperfect electronic spin reinitialization in $\ket{0}$ might lead to decoherence.

To investigate the effect of the former, we repeat the measurement of the DPS with the smallest dephasing, formed by nuclear spins 2 and 3, but this time with single MW pulses rather than the BB1 pulses used in the main text. Albeit this leads to a reduced MW pulse fidelity, we observe no decrease of the decay constant. In addition, we perform a measurement in which we replace the MW $\pi/2$ pulse of the entanglement sequence by a $\pi$ pulse, such that the electronic spin ends up in $\ket{0}$ at the end of the sequence and no repumping is required. We observe that the dephasing now follows a Gaussian decay and is strongly reduced compared to the case of an initial $\pi/2$ pulse. The observed decay can be fully explained by the independently measured nuclear spin dephasing time $T_2^*$, which testifies that the electronic spin reset is the dominant dephasing mechanism in our experiment.

Effects from imperfect reinitialization in $\ket{0}$ seem unlikely from the remaining population in $\ket{\pm1}$, $p_{\pm1}<0.2(2)\,\%$, measured after optical pumping for $1.5\,\mu$s using $2\,\mu$W of laser power. This expectation is confirmed by the observation of an unchanged decay constant upon stepwise increasing the duration of the repumping laser pulse up to $5.5\,\mu$s. Similarly, an additional delay of several $\mu\text{s}$ after the repumping, which would increase the time spent in a state that causes dephasing whenever the pumping process is not perfect, has no discernible effect.

We therefore conclude that the dephasing caused by experimental imperfections is negligible in the measurements presented in the main text.

\subsection{Power dependence of the optical pumping timescales and NV deionization probability}
\begin{figure}
\includegraphics[width=\columnwidth]{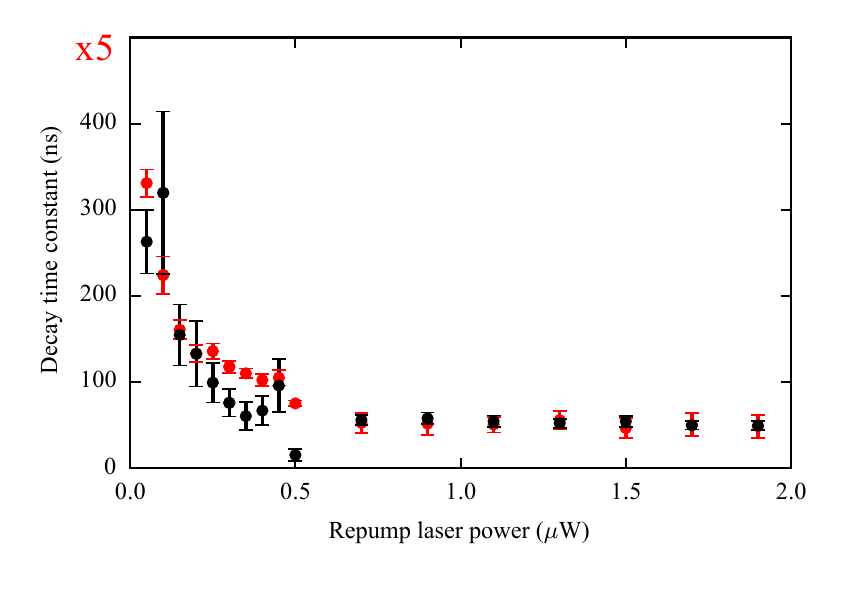}
\includegraphics[width=\columnwidth]{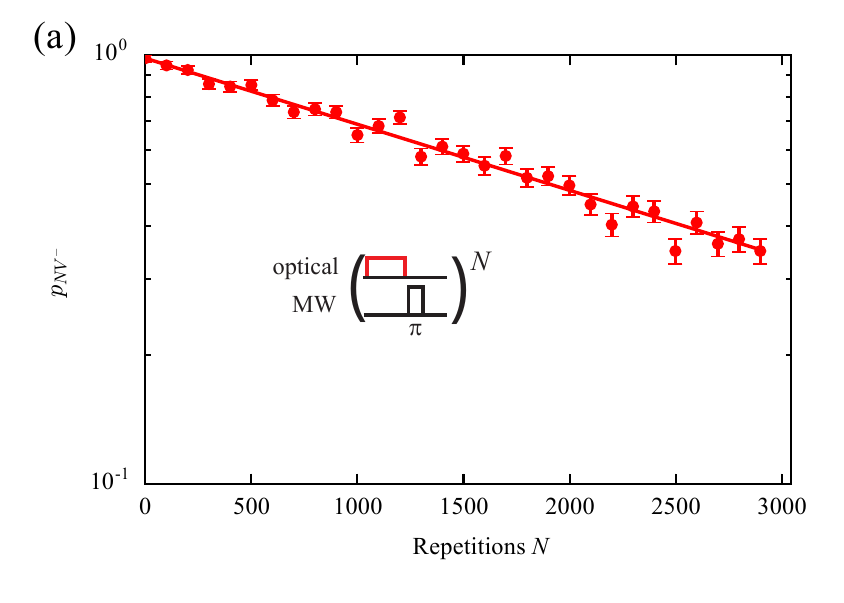}
\caption{ \label{fig:reset_ionization}
(a) Power dependence of the NV electronic spin reset timescales (black: short, red: long) using $2\,\mu$W in the $\ket{E_{1,2}}$ repump configuration.
(b) Fraction of experimental runs in which the NV center is not deionized to the neutral charge state as a function of the number of electron resets when applying 2 $\mu$W of laser power in the sequence shown as inset. We observe an exponential decay with a decay constant $2.82(8)\times 10^{3}$.
}
\end{figure}

In the main text, we demonstrate that a fast reset of the electronic spin is required to minimize decoherence of the coupled nuclear spins. The reset is doubly exponential, with timescales that depend on the applied laser power. Fig. \ref{fig:reset_ionization}(a) shows the power dependence of these timescales measured in the $\ket{A_{1,2}}$ repump laser configuration. We observe a saturation at $\sim 500$ nW of applied laser power. To ensure the fastest possible reset even in the case of NV position or laser power drift, we use $2\,\mu$W, which is well above saturation, in all measurements presented in the main text except where mentioned otherwise.

Such comparably high repump power has the unwanted effect that the NV can be transferred to the neutral charge state, which leads to a detectable (and thus heralded) loss of the nuclear spin qubit. This deionization is likely a two-photon process \cite{doherty_nitrogen-vacancy_2013}, such that we expect its probability to increase quadratically with applied laser power.

To investigate the effects of varying reset timescales and hyperfine coupling strengths on the nuclear spin dephasing without added noise, we post-select the analysis of the results presented in the main text to experimental runs in which the NV has not been deionized. To this end, we measure the NV charge state after completion of the experimental sequence by applying resonant lasers on both the cycling and repump transitions for $50\,\mu$s. If the NV is in the negative charge state, we detect on average more than 10 photons in this time interval. If we detect less than two photons, the NV has likely been deionized.

To characterize the influence of NV deionization on the results presented in the main text, we measure the deionization probability as a function of the number of electron resets when applying 2 $\mu$W of laser power in the $\ket{A_{1,2}}$ repump configuration, i.e. we study the highest powers used in the main text. Note that in this experiment the electron is prepared in $\ket{-1}$ and thus reset in \emph{every} repetition. As can be seen in Fig. \ref{fig:reset_ionization}(b), we observe an exponential decay of the probability to remain in the negatively charged state, $p_{NV^-}$, with a decay constant of $N_{d}=2.82(8)\times 10^3$ repetitions. In the entanglement sequence, where repumping is required in $50\,\%$ of all trials (on average), we expect the decay to be two times slower. Albeit the resulting decay is larger than all observed dephasing constants, it becomes evident that a further reduction of the induced decoherence (e.g. via the use of decoherence-protected subspaces with even smaller coupling strengths than the ones investigated in the main text) will likely require a reduction of the maximally applied laser power.

\subsection{Nuclear spin polarization decay and natural dephasing correction}
\begin{figure}
\includegraphics[width=\columnwidth]{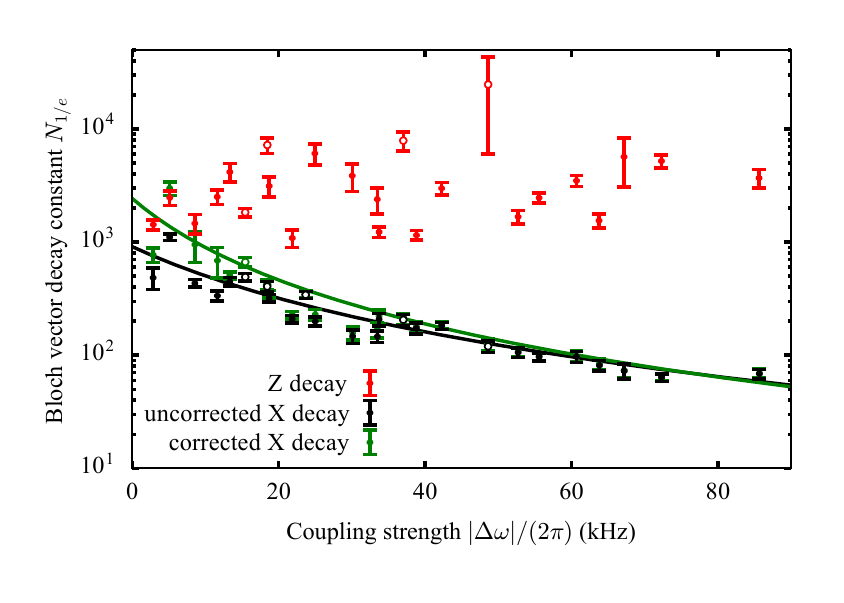}
\caption{ \label{fig:Z_decay}
Decay of the nuclear spin Bloch vector length with increasing coupling strength for $2\,\mu$W repump laser power applied in the $\ket{A_{1,2}}$ configuration (see main text). The open circles are single nuclear spins, while the filled circles are two-spin subspaces. The red data shows the population decay of a qubit that is encoded in an eigenstate, i.e. along the axis of the external magnetic field. This decay is typically much smaller than the dephasing of a qubit prepared in a superposition state (black and green). The black and green curve show the same measurements, analyzed either with (green) or without (black) applying a correction for the intrinsic nuclear spin dephasing time $T_2^*$ (see text). A significant deviation is only observed for the leftmost data points.
}
\end{figure}

As explained in the main text, we expect that the dominant decoherence mechanism of the nuclear spins when applying the entanglement sequence is dephasing which is induced by the electronic spin reset. In contrast, the nuclear spin population is expected to be much less sensitive. To investigate this, we initialize a qubit in the eigenstates $\ket{\downarrow}$ for each single nuclear spin and in $\ket{\downarrow\downarrow}$ and $\ket{\uparrow\downarrow}$ for each investigated two-spin configuration. We then repeat the entanglement sequence many times and measure the exponential decay of the qubit $Z$ projection. As can be seen in Fig.
\ref{fig:Z_decay}, the population decay (red) is typically much smaller than the dephasing shown in black (data identical to Fig. 4(b) of the main text).

In addition to the decoherence that is induced by repeatedly applying the entanglement sequence, the nuclear spins also exhibit natural dephasing. In Ref. \cite{cramer_repeated_2015}, this dephasing was shown to lead to a Gaussian decay of the state fidelity with a timescale $T_2^*$ that is different for each measured $\C$ spin, c.f. Table I of the main text. This dephasing can be mitigated by a suited dynamical decoupling sequence on the nuclear spins, which can e.g. be implemented by using radio-frequency pulses that are resonant to the nuclear spin Larmor frequency \cite{maurer_room-temperature_2012}.

In our experiments, we do not implement such nuclear spin decoupling sequence. Instead, before fitting the decay we correct the obtained data points for their natural dephasing. To this end, we divide by the factor $\exp(-(\frac{t_N}{T_2^*})^2)$, where $t_N$ is the absolute time elapsed after doing $N$ repetitions of the entanglement sequence and $T_2^*$ is the effective natural dephasing time. For single nuclear spins, we take the measured value of $T_2^*$, whereas we extrapolate the value for a DPS that is formed by the spins $i$ and $j$ as $T_{2,ij}^* \approx \frac{1}{\sqrt{(1/T^*_{2,i})^2+(1/T^*_{2,j})^2}}$. In Fig. \ref{fig:Z_decay}, we show both the corrected (green) and uncorrected (black) data, where it becomes evident that natural dephasing only becomes relevant for DPSs with very low effective coupling strengths.

\begin{acknowledgments}
We thank Adam Gali for discussions on the NV singlet states. We acknowledge support from the Dutch Organization for Fundamental Research on Matter (FOM), the Netherlands Organization for Scientic Research (NWO) through a VENI grant (THT), and the European Research Council through a Starting Grant (RH).
\end{acknowledgments}

\newpage
\bibliography{Library}

\begin{thebibliography}{43}%
\makeatletter
\providecommand \@ifxundefined [1]{%
 \@ifx{#1\undefined}
}%
\providecommand \@ifnum [1]{%
 \ifnum #1\expandafter \@firstoftwo
 \else \expandafter \@secondoftwo
 \fi
}%
\providecommand \@ifx [1]{%
 \ifx #1\expandafter \@firstoftwo
 \else \expandafter \@secondoftwo
 \fi
}%
\providecommand \natexlab [1]{#1}%
\providecommand \enquote  [1]{``#1''}%
\providecommand \bibnamefont  [1]{#1}%
\providecommand \bibfnamefont [1]{#1}%
\providecommand \citenamefont [1]{#1}%
\providecommand \href@noop [0]{\@secondoftwo}%
\providecommand \href [0]{\begingroup \@sanitize@url \@href}%
\providecommand \@href[1]{\@@startlink{#1}\@@href}%
\providecommand \@@href[1]{\endgroup#1\@@endlink}%
\providecommand \@sanitize@url [0]{\catcode `\\12\catcode `\$12\catcode
  `\&12\catcode `\#12\catcode `\^12\catcode `\_12\catcode `\%12\relax}%
\providecommand \@@startlink[1]{}%
\providecommand \@@endlink[0]{}%
\providecommand \url  [0]{\begingroup\@sanitize@url \@url }%
\providecommand \@url [1]{\endgroup\@href {#1}{\urlprefix }}%
\providecommand \urlprefix  [0]{URL }%
\providecommand \Eprint [0]{\href }%
\providecommand \doibase [0]{http://dx.doi.org/}%
\providecommand \selectlanguage [0]{\@gobble}%
\providecommand \bibinfo  [0]{\@secondoftwo}%
\providecommand \bibfield  [0]{\@secondoftwo}%
\providecommand \translation [1]{[#1]}%
\providecommand \BibitemOpen [0]{}%
\providecommand \bibitemStop [0]{}%
\providecommand \bibitemNoStop [0]{.\EOS\space}%
\providecommand \EOS [0]{\spacefactor3000\relax}%
\providecommand \BibitemShut  [1]{\csname bibitem#1\endcsname}%
\let\auto@bib@innerbib\@empty
\bibitem [{\citenamefont {Kimble}(2008)}]{kimble_quantum_2008}%
  \BibitemOpen
  \bibfield  {author} {\bibinfo {author} {\bibfnamefont {H.~J.}\ \bibnamefont
  {Kimble}},\ }\href {\doibase 10.1038/nature07127} {\bibfield  {journal}
  {\bibinfo  {journal} {Nature}\ }\textbf {\bibinfo {volume} {453}},\ \bibinfo
  {pages} {1023} (\bibinfo {year} {2008})}\BibitemShut {NoStop}%
\bibitem [{\citenamefont {Duan}\ and\ \citenamefont
  {Monroe}(2010)}]{duan_colloquium:_2010}%
  \BibitemOpen
  \bibfield  {author} {\bibinfo {author} {\bibfnamefont {L.-M.}\ \bibnamefont
  {Duan}}\ and\ \bibinfo {author} {\bibfnamefont {C.}~\bibnamefont {Monroe}},\
  }\href {\doibase 10.1103/RevModPhys.82.1209} {\bibfield  {journal} {\bibinfo
  {journal} {Rev. Mod. Phys.}\ }\textbf {\bibinfo {volume} {82}},\ \bibinfo
  {pages} {1209} (\bibinfo {year} {2010})}\BibitemShut {NoStop}%
\bibitem [{\citenamefont {Sangouard}\ \emph {et~al.}(2011)\citenamefont
  {Sangouard}, \citenamefont {Simon}, \citenamefont {de~Riedmatten},\ and\
  \citenamefont {Gisin}}]{sangouard_quantum_2011}%
  \BibitemOpen
  \bibfield  {author} {\bibinfo {author} {\bibfnamefont {N.}~\bibnamefont
  {Sangouard}}, \bibinfo {author} {\bibfnamefont {C.}~\bibnamefont {Simon}},
  \bibinfo {author} {\bibfnamefont {H.}~\bibnamefont {de~Riedmatten}}, \ and\
  \bibinfo {author} {\bibfnamefont {N.}~\bibnamefont {Gisin}},\ }\href
  {\doibase 10.1103/RevModPhys.83.33} {\bibfield  {journal} {\bibinfo
  {journal} {Rev. Mod. Phys.}\ }\textbf {\bibinfo {volume} {83}},\ \bibinfo
  {pages} {33} (\bibinfo {year} {2011})}\BibitemShut {NoStop}%
\bibitem [{\citenamefont {Reiserer}\ and\ \citenamefont
  {Rempe}(2015)}]{reiserer_cavity-based_2015}%
  \BibitemOpen
  \bibfield  {author} {\bibinfo {author} {\bibfnamefont {A.}~\bibnamefont
  {Reiserer}}\ and\ \bibinfo {author} {\bibfnamefont {G.}~\bibnamefont
  {Rempe}},\ }\href {\doibase 10.1103/RevModPhys.87.1379} {\bibfield  {journal}
  {\bibinfo  {journal} {Rev. Mod. Phys.}\ }\textbf {\bibinfo {volume} {87}},\
  \bibinfo {pages} {1379} (\bibinfo {year} {2015})}\BibitemShut {NoStop}%
\bibitem [{\citenamefont {Bancal}\ \emph {et~al.}(2012)\citenamefont {Bancal},
  \citenamefont {Pironio}, \citenamefont {Ac{\'i}n}, \citenamefont {Liang},
  \citenamefont {Scarani},\ and\ \citenamefont {Gisin}}]{bancal_quantum_2012}%
  \BibitemOpen
  \bibfield  {author} {\bibinfo {author} {\bibfnamefont {J.-D.}\ \bibnamefont
  {Bancal}}, \bibinfo {author} {\bibfnamefont {S.}~\bibnamefont {Pironio}},
  \bibinfo {author} {\bibfnamefont {A.}~\bibnamefont {Ac{\'i}n}}, \bibinfo
  {author} {\bibfnamefont {Y.-C.}\ \bibnamefont {Liang}}, \bibinfo {author}
  {\bibfnamefont {V.}~\bibnamefont {Scarani}}, \ and\ \bibinfo {author}
  {\bibfnamefont {N.}~\bibnamefont {Gisin}},\ }\href {\doibase
  10.1038/nphys2460} {\bibfield  {journal} {\bibinfo  {journal} {Nat. Phys.}\
  }\textbf {\bibinfo {volume} {8}},\ \bibinfo {pages} {867} (\bibinfo {year}
  {2012})}\BibitemShut {NoStop}%
\bibitem [{\citenamefont {K{\'o}m{\'a}r}\ \emph {et~al.}(2014)\citenamefont
  {K{\'o}m{\'a}r}, \citenamefont {Kessler}, \citenamefont {Bishof},
  \citenamefont {Jiang}, \citenamefont {S{\o}rensen}, \citenamefont {Ye},\ and\
  \citenamefont {Lukin}}]{komar_quantum_2014}%
  \BibitemOpen
  \bibfield  {author} {\bibinfo {author} {\bibfnamefont {P.}~\bibnamefont
  {K{\'o}m{\'a}r}}, \bibinfo {author} {\bibfnamefont {E.~M.}\ \bibnamefont
  {Kessler}}, \bibinfo {author} {\bibfnamefont {M.}~\bibnamefont {Bishof}},
  \bibinfo {author} {\bibfnamefont {L.}~\bibnamefont {Jiang}}, \bibinfo
  {author} {\bibfnamefont {A.~S.}\ \bibnamefont {S{\o}rensen}}, \bibinfo
  {author} {\bibfnamefont {J.}~\bibnamefont {Ye}}, \ and\ \bibinfo {author}
  {\bibfnamefont {M.~D.}\ \bibnamefont {Lukin}},\ }\href {\doibase
  10.1038/nphys3000} {\bibfield  {journal} {\bibinfo  {journal} {Nat. Phys.}\
  }\textbf {\bibinfo {volume} {10}},\ \bibinfo {pages} {582} (\bibinfo {year}
  {2014})}\BibitemShut {NoStop}%
\bibitem [{\citenamefont {Nickerson}\ \emph {et~al.}(2013)\citenamefont
  {Nickerson}, \citenamefont {Li},\ and\ \citenamefont
  {Benjamin}}]{nickerson_topological_2013}%
  \BibitemOpen
  \bibfield  {author} {\bibinfo {author} {\bibfnamefont {N.~H.}\ \bibnamefont
  {Nickerson}}, \bibinfo {author} {\bibfnamefont {Y.}~\bibnamefont {Li}}, \
  and\ \bibinfo {author} {\bibfnamefont {S.~C.}\ \bibnamefont {Benjamin}},\
  }\href {\doibase 10.1038/ncomms2773} {\bibfield  {journal} {\bibinfo
  {journal} {Nat. Commun.}\ }\textbf {\bibinfo {volume} {4}},\ \bibinfo {pages}
  {1756} (\bibinfo {year} {2013})}\BibitemShut {NoStop}%
\bibitem [{\citenamefont {Barz}\ \emph {et~al.}(2012)\citenamefont {Barz},
  \citenamefont {Kashefi}, \citenamefont {Broadbent}, \citenamefont
  {Fitzsimons}, \citenamefont {Zeilinger},\ and\ \citenamefont
  {Walther}}]{barz_demonstration_2012}%
  \BibitemOpen
  \bibfield  {author} {\bibinfo {author} {\bibfnamefont {S.}~\bibnamefont
  {Barz}}, \bibinfo {author} {\bibfnamefont {E.}~\bibnamefont {Kashefi}},
  \bibinfo {author} {\bibfnamefont {A.}~\bibnamefont {Broadbent}}, \bibinfo
  {author} {\bibfnamefont {J.~F.}\ \bibnamefont {Fitzsimons}}, \bibinfo
  {author} {\bibfnamefont {A.}~\bibnamefont {Zeilinger}}, \ and\ \bibinfo
  {author} {\bibfnamefont {P.}~\bibnamefont {Walther}},\ }\href {\doibase
  10.1126/science.1214707} {\bibfield  {journal} {\bibinfo  {journal}
  {Science}\ }\textbf {\bibinfo {volume} {335}},\ \bibinfo {pages} {303}
  (\bibinfo {year} {2012})}\BibitemShut {NoStop}%
\bibitem [{\citenamefont {Ekert}\ and\ \citenamefont
  {Renner}(2014)}]{ekert_ultimate_2014}%
  \BibitemOpen
  \bibfield  {author} {\bibinfo {author} {\bibfnamefont {A.}~\bibnamefont
  {Ekert}}\ and\ \bibinfo {author} {\bibfnamefont {R.}~\bibnamefont {Renner}},\
  }\href {\doibase 10.1038/nature13132} {\bibfield  {journal} {\bibinfo
  {journal} {Nature}\ }\textbf {\bibinfo {volume} {507}},\ \bibinfo {pages}
  {443} (\bibinfo {year} {2014})}\BibitemShut {NoStop}%
\bibitem [{\citenamefont {Hofmann}\ \emph {et~al.}(2012)\citenamefont
  {Hofmann}, \citenamefont {Krug}, \citenamefont {Ortegel}, \citenamefont
  {G{\'e}rard}, \citenamefont {Weber}, \citenamefont {Rosenfeld},\ and\
  \citenamefont {Weinfurter}}]{hofmann_heralded_2012}%
  \BibitemOpen
  \bibfield  {author} {\bibinfo {author} {\bibfnamefont {J.}~\bibnamefont
  {Hofmann}}, \bibinfo {author} {\bibfnamefont {M.}~\bibnamefont {Krug}},
  \bibinfo {author} {\bibfnamefont {N.}~\bibnamefont {Ortegel}}, \bibinfo
  {author} {\bibfnamefont {L.}~\bibnamefont {G{\'e}rard}}, \bibinfo {author}
  {\bibfnamefont {M.}~\bibnamefont {Weber}}, \bibinfo {author} {\bibfnamefont
  {W.}~\bibnamefont {Rosenfeld}}, \ and\ \bibinfo {author} {\bibfnamefont
  {H.}~\bibnamefont {Weinfurter}},\ }\href {\doibase 10.1126/science.1221856}
  {\bibfield  {journal} {\bibinfo  {journal} {Science}\ }\textbf {\bibinfo
  {volume} {337}},\ \bibinfo {pages} {72} (\bibinfo {year} {2012})}\BibitemShut
  {NoStop}%
\bibitem [{\citenamefont {Hucul}\ \emph {et~al.}(2015)\citenamefont {Hucul},
  \citenamefont {Inlek}, \citenamefont {Vittorini}, \citenamefont {Crocker},
  \citenamefont {Debnath}, \citenamefont {Clark},\ and\ \citenamefont
  {Monroe}}]{hucul_modular_2015}%
  \BibitemOpen
  \bibfield  {author} {\bibinfo {author} {\bibfnamefont {D.}~\bibnamefont
  {Hucul}}, \bibinfo {author} {\bibfnamefont {I.~V.}\ \bibnamefont {Inlek}},
  \bibinfo {author} {\bibfnamefont {G.}~\bibnamefont {Vittorini}}, \bibinfo
  {author} {\bibfnamefont {C.}~\bibnamefont {Crocker}}, \bibinfo {author}
  {\bibfnamefont {S.}~\bibnamefont {Debnath}}, \bibinfo {author} {\bibfnamefont
  {S.~M.}\ \bibnamefont {Clark}}, \ and\ \bibinfo {author} {\bibfnamefont
  {C.}~\bibnamefont {Monroe}},\ }\href {\doibase 10.1038/nphys3150} {\bibfield
  {journal} {\bibinfo  {journal} {Nat. Phys.}\ }\textbf {\bibinfo {volume}
  {11}},\ \bibinfo {pages} {37} (\bibinfo {year} {2015})}\BibitemShut {NoStop}%
\bibitem [{\citenamefont {Gao}\ \emph {et~al.}(2015)\citenamefont {Gao},
  \citenamefont {Imamoglu}, \citenamefont {Bernien},\ and\ \citenamefont
  {Hanson}}]{gao_coherent_2015}%
  \BibitemOpen
  \bibfield  {author} {\bibinfo {author} {\bibfnamefont {W.~B.}\ \bibnamefont
  {Gao}}, \bibinfo {author} {\bibfnamefont {A.}~\bibnamefont {Imamoglu}},
  \bibinfo {author} {\bibfnamefont {H.}~\bibnamefont {Bernien}}, \ and\
  \bibinfo {author} {\bibfnamefont {R.}~\bibnamefont {Hanson}},\ }\href
  {\doibase 10.1038/nphoton.2015.58} {\bibfield  {journal} {\bibinfo  {journal}
  {Nat. Photon.}\ }\textbf {\bibinfo {volume} {9}},\ \bibinfo {pages} {363}
  (\bibinfo {year} {2015})}\BibitemShut {NoStop}%
\bibitem [{\citenamefont {Hensen}\ \emph {et~al.}(2015)\citenamefont {Hensen},
  \citenamefont {Bernien}, \citenamefont {Dr{\'e}au}, \citenamefont {Reiserer},
  \citenamefont {Kalb}, \citenamefont {Blok}, \citenamefont {Ruitenberg},
  \citenamefont {Vermeulen}, \citenamefont {Schouten}, \citenamefont
  {Abell{\'a}n}, \citenamefont {Amaya}, \citenamefont {Pruneri}, \citenamefont
  {Mitchell}, \citenamefont {Markham}, \citenamefont {Twitchen}, \citenamefont
  {Elkouss}, \citenamefont {Wehner}, \citenamefont {Taminiau},\ and\
  \citenamefont {Hanson}}]{hensen_loophole-free_2015}%
  \BibitemOpen
  \bibfield  {author} {\bibinfo {author} {\bibfnamefont {B.}~\bibnamefont
  {Hensen}}, \bibinfo {author} {\bibfnamefont {H.}~\bibnamefont {Bernien}},
  \bibinfo {author} {\bibfnamefont {A.~E.}\ \bibnamefont {Dr{\'e}au}}, \bibinfo
  {author} {\bibfnamefont {A.}~\bibnamefont {Reiserer}}, \bibinfo {author}
  {\bibfnamefont {N.}~\bibnamefont {Kalb}}, \bibinfo {author} {\bibfnamefont
  {M.~S.}\ \bibnamefont {Blok}}, \bibinfo {author} {\bibfnamefont
  {J.}~\bibnamefont {Ruitenberg}}, \bibinfo {author} {\bibfnamefont {R.~F.~L.}\
  \bibnamefont {Vermeulen}}, \bibinfo {author} {\bibfnamefont {R.~N.}\
  \bibnamefont {Schouten}}, \bibinfo {author} {\bibfnamefont {C.}~\bibnamefont
  {Abell{\'a}n}}, \bibinfo {author} {\bibfnamefont {W.}~\bibnamefont {Amaya}},
  \bibinfo {author} {\bibfnamefont {V.}~\bibnamefont {Pruneri}}, \bibinfo
  {author} {\bibfnamefont {M.~W.}\ \bibnamefont {Mitchell}}, \bibinfo {author}
  {\bibfnamefont {M.}~\bibnamefont {Markham}}, \bibinfo {author} {\bibfnamefont
  {D.~J.}\ \bibnamefont {Twitchen}}, \bibinfo {author} {\bibfnamefont
  {D.}~\bibnamefont {Elkouss}}, \bibinfo {author} {\bibfnamefont
  {S.}~\bibnamefont {Wehner}}, \bibinfo {author} {\bibfnamefont {T.~H.}\
  \bibnamefont {Taminiau}}, \ and\ \bibinfo {author} {\bibfnamefont
  {R.}~\bibnamefont {Hanson}},\ }\href {\doibase 10.1038/nature15759}
  {\bibfield  {journal} {\bibinfo  {journal} {Nature}\ }\textbf {\bibinfo
  {volume} {526}},\ \bibinfo {pages} {682} (\bibinfo {year}
  {2015})}\BibitemShut {NoStop}%
\bibitem [{\citenamefont {Delteil}\ \emph {et~al.}(2015)\citenamefont
  {Delteil}, \citenamefont {Sun}, \citenamefont {Gao}, \citenamefont {Togan},
  \citenamefont {Faelt},\ and\ \citenamefont {Imamo{\u
  g}lu}}]{delteil_generation_2015}%
  \BibitemOpen
  \bibfield  {author} {\bibinfo {author} {\bibfnamefont {A.}~\bibnamefont
  {Delteil}}, \bibinfo {author} {\bibfnamefont {Z.}~\bibnamefont {Sun}},
  \bibinfo {author} {\bibfnamefont {W.-b.}\ \bibnamefont {Gao}}, \bibinfo
  {author} {\bibfnamefont {E.}~\bibnamefont {Togan}}, \bibinfo {author}
  {\bibfnamefont {S.}~\bibnamefont {Faelt}}, \ and\ \bibinfo {author}
  {\bibfnamefont {A.}~\bibnamefont {Imamo{\u g}lu}},\ }\href {\doibase
  10.1038/nphys3605} {\bibfield  {journal} {\bibinfo  {journal} {Nat. Phys.}\ }
  (\bibinfo {year} {2015}),\ 10.1038/nphys3605}\BibitemShut {NoStop}%
\bibitem [{\citenamefont {Bennett}\ \emph {et~al.}(1996)\citenamefont
  {Bennett}, \citenamefont {Brassard}, \citenamefont {Popescu}, \citenamefont
  {Schumacher}, \citenamefont {Smolin},\ and\ \citenamefont
  {Wootters}}]{bennett_purification_1996}%
  \BibitemOpen
  \bibfield  {author} {\bibinfo {author} {\bibfnamefont {C.~H.}\ \bibnamefont
  {Bennett}}, \bibinfo {author} {\bibfnamefont {G.}~\bibnamefont {Brassard}},
  \bibinfo {author} {\bibfnamefont {S.}~\bibnamefont {Popescu}}, \bibinfo
  {author} {\bibfnamefont {B.}~\bibnamefont {Schumacher}}, \bibinfo {author}
  {\bibfnamefont {J.~A.}\ \bibnamefont {Smolin}}, \ and\ \bibinfo {author}
  {\bibfnamefont {W.~K.}\ \bibnamefont {Wootters}},\ }\href {\doibase
  10.1103/PhysRevLett.76.722} {\bibfield  {journal} {\bibinfo  {journal} {Phys.
  Rev. Lett.}\ }\textbf {\bibinfo {volume} {76}},\ \bibinfo {pages} {722}
  (\bibinfo {year} {1996})}\BibitemShut {NoStop}%
\bibitem [{\citenamefont {Briegel}\ \emph {et~al.}(1998)\citenamefont
  {Briegel}, \citenamefont {D{\"u}r}, \citenamefont {Cirac},\ and\
  \citenamefont {Zoller}}]{briegel_quantum_1998}%
  \BibitemOpen
  \bibfield  {author} {\bibinfo {author} {\bibfnamefont {H.-J.}\ \bibnamefont
  {Briegel}}, \bibinfo {author} {\bibfnamefont {W.}~\bibnamefont {D{\"u}r}},
  \bibinfo {author} {\bibfnamefont {J.~I.}\ \bibnamefont {Cirac}}, \ and\
  \bibinfo {author} {\bibfnamefont {P.}~\bibnamefont {Zoller}},\ }\href
  {\doibase 10.1103/PhysRevLett.81.5932} {\bibfield  {journal} {\bibinfo
  {journal} {Phys. Rev. Lett.}\ }\textbf {\bibinfo {volume} {81}},\ \bibinfo
  {pages} {5932} (\bibinfo {year} {1998})}\BibitemShut {NoStop}%
\bibitem [{\citenamefont {Childress}\ \emph {et~al.}(2006)\citenamefont
  {Childress}, \citenamefont {Taylor}, \citenamefont {S{\o}rensen},\ and\
  \citenamefont {Lukin}}]{childress_fault-tolerant_2006}%
  \BibitemOpen
  \bibfield  {author} {\bibinfo {author} {\bibfnamefont {L.}~\bibnamefont
  {Childress}}, \bibinfo {author} {\bibfnamefont {J.~M.}\ \bibnamefont
  {Taylor}}, \bibinfo {author} {\bibfnamefont {A.~S.}\ \bibnamefont
  {S{\o}rensen}}, \ and\ \bibinfo {author} {\bibfnamefont {M.~D.}\ \bibnamefont
  {Lukin}},\ }\href {\doibase 10.1103/PhysRevLett.96.070504} {\bibfield
  {journal} {\bibinfo  {journal} {Phys. Rev. Lett.}\ }\textbf {\bibinfo
  {volume} {96}},\ \bibinfo {pages} {070504} (\bibinfo {year}
  {2006})}\BibitemShut {NoStop}%
\bibitem [{\citenamefont {Fowler}\ \emph {et~al.}(2010)\citenamefont {Fowler},
  \citenamefont {Wang}, \citenamefont {Hill}, \citenamefont {Ladd},
  \citenamefont {Van~Meter},\ and\ \citenamefont
  {Hollenberg}}]{fowler_surface_2010}%
  \BibitemOpen
  \bibfield  {author} {\bibinfo {author} {\bibfnamefont {A.~G.}\ \bibnamefont
  {Fowler}}, \bibinfo {author} {\bibfnamefont {D.~S.}\ \bibnamefont {Wang}},
  \bibinfo {author} {\bibfnamefont {C.~D.}\ \bibnamefont {Hill}}, \bibinfo
  {author} {\bibfnamefont {T.~D.}\ \bibnamefont {Ladd}}, \bibinfo {author}
  {\bibfnamefont {R.}~\bibnamefont {Van~Meter}}, \ and\ \bibinfo {author}
  {\bibfnamefont {L.~C.~L.}\ \bibnamefont {Hollenberg}},\ }\href {\doibase
  10.1103/PhysRevLett.104.180503} {\bibfield  {journal} {\bibinfo  {journal}
  {Phys. Rev. Lett.}\ }\textbf {\bibinfo {volume} {104}},\ \bibinfo {pages}
  {180503} (\bibinfo {year} {2010})}\BibitemShut {NoStop}%
\bibitem [{\citenamefont {Duan}\ \emph {et~al.}(2001)\citenamefont {Duan},
  \citenamefont {Lukin}, \citenamefont {Cirac},\ and\ \citenamefont
  {Zoller}}]{duan_long-distance_2001}%
  \BibitemOpen
  \bibfield  {author} {\bibinfo {author} {\bibfnamefont {L.-M.}\ \bibnamefont
  {Duan}}, \bibinfo {author} {\bibfnamefont {M.~D.}\ \bibnamefont {Lukin}},
  \bibinfo {author} {\bibfnamefont {J.~I.}\ \bibnamefont {Cirac}}, \ and\
  \bibinfo {author} {\bibfnamefont {P.}~\bibnamefont {Zoller}},\ }\href
  {\doibase 10.1038/35106500} {\bibfield  {journal} {\bibinfo  {journal}
  {Nature}\ }\textbf {\bibinfo {volume} {414}},\ \bibinfo {pages} {413}
  (\bibinfo {year} {2001})}\BibitemShut {NoStop}%
\bibitem [{\citenamefont {Bernien}\ \emph {et~al.}(2013)\citenamefont
  {Bernien}, \citenamefont {Hensen}, \citenamefont {Pfaff}, \citenamefont
  {Koolstra}, \citenamefont {Blok}, \citenamefont {Robledo}, \citenamefont
  {Taminiau}, \citenamefont {Markham}, \citenamefont {Twitchen}, \citenamefont
  {Childress},\ and\ \citenamefont {Hanson}}]{bernien_heralded_2013}%
  \BibitemOpen
  \bibfield  {author} {\bibinfo {author} {\bibfnamefont {H.}~\bibnamefont
  {Bernien}}, \bibinfo {author} {\bibfnamefont {B.}~\bibnamefont {Hensen}},
  \bibinfo {author} {\bibfnamefont {W.}~\bibnamefont {Pfaff}}, \bibinfo
  {author} {\bibfnamefont {G.}~\bibnamefont {Koolstra}}, \bibinfo {author}
  {\bibfnamefont {M.~S.}\ \bibnamefont {Blok}}, \bibinfo {author}
  {\bibfnamefont {L.}~\bibnamefont {Robledo}}, \bibinfo {author} {\bibfnamefont
  {T.~H.}\ \bibnamefont {Taminiau}}, \bibinfo {author} {\bibfnamefont
  {M.}~\bibnamefont {Markham}}, \bibinfo {author} {\bibfnamefont {D.~J.}\
  \bibnamefont {Twitchen}}, \bibinfo {author} {\bibfnamefont {L.}~\bibnamefont
  {Childress}}, \ and\ \bibinfo {author} {\bibfnamefont {R.}~\bibnamefont
  {Hanson}},\ }\href {\doibase 10.1038/nature12016} {\bibfield  {journal}
  {\bibinfo  {journal} {Nature}\ }\textbf {\bibinfo {volume} {497}},\ \bibinfo
  {pages} {86} (\bibinfo {year} {2013})}\BibitemShut {NoStop}%
\bibitem [{\citenamefont {Pfaff}\ \emph {et~al.}(2014)\citenamefont {Pfaff},
  \citenamefont {Hensen}, \citenamefont {Bernien}, \citenamefont {Dam},
  \citenamefont {Blok}, \citenamefont {Taminiau}, \citenamefont {Tiggelman},
  \citenamefont {Schouten}, \citenamefont {Markham}, \citenamefont {Twitchen},\
  and\ \citenamefont {Hanson}}]{pfaff_unconditional_2014}%
  \BibitemOpen
  \bibfield  {author} {\bibinfo {author} {\bibfnamefont {W.}~\bibnamefont
  {Pfaff}}, \bibinfo {author} {\bibfnamefont {B.~J.}\ \bibnamefont {Hensen}},
  \bibinfo {author} {\bibfnamefont {H.}~\bibnamefont {Bernien}}, \bibinfo
  {author} {\bibfnamefont {S.~B.~v.}\ \bibnamefont {Dam}}, \bibinfo {author}
  {\bibfnamefont {M.~S.}\ \bibnamefont {Blok}}, \bibinfo {author}
  {\bibfnamefont {T.~H.}\ \bibnamefont {Taminiau}}, \bibinfo {author}
  {\bibfnamefont {M.~J.}\ \bibnamefont {Tiggelman}}, \bibinfo {author}
  {\bibfnamefont {R.~N.}\ \bibnamefont {Schouten}}, \bibinfo {author}
  {\bibfnamefont {M.}~\bibnamefont {Markham}}, \bibinfo {author} {\bibfnamefont
  {D.~J.}\ \bibnamefont {Twitchen}}, \ and\ \bibinfo {author} {\bibfnamefont
  {R.}~\bibnamefont {Hanson}},\ }\href {\doibase 10.1126/science.1253512}
  {\bibfield  {journal} {\bibinfo  {journal} {Science}\ }\textbf {\bibinfo
  {volume} {345}},\ \bibinfo {pages} {532} (\bibinfo {year}
  {2014})}\BibitemShut {NoStop}%
\bibitem [{\citenamefont {Dutt}\ \emph {et~al.}(2007)\citenamefont {Dutt},
  \citenamefont {Childress}, \citenamefont {Jiang}, \citenamefont {Togan},
  \citenamefont {Maze}, \citenamefont {Jelezko}, \citenamefont {Zibrov},
  \citenamefont {Hemmer},\ and\ \citenamefont {Lukin}}]{dutt_quantum_2007}%
  \BibitemOpen
  \bibfield  {author} {\bibinfo {author} {\bibfnamefont {M.~V.~G.}\
  \bibnamefont {Dutt}}, \bibinfo {author} {\bibfnamefont {L.}~\bibnamefont
  {Childress}}, \bibinfo {author} {\bibfnamefont {L.}~\bibnamefont {Jiang}},
  \bibinfo {author} {\bibfnamefont {E.}~\bibnamefont {Togan}}, \bibinfo
  {author} {\bibfnamefont {J.}~\bibnamefont {Maze}}, \bibinfo {author}
  {\bibfnamefont {F.}~\bibnamefont {Jelezko}}, \bibinfo {author} {\bibfnamefont
  {A.~S.}\ \bibnamefont {Zibrov}}, \bibinfo {author} {\bibfnamefont {P.~R.}\
  \bibnamefont {Hemmer}}, \ and\ \bibinfo {author} {\bibfnamefont {M.~D.}\
  \bibnamefont {Lukin}},\ }\href {\doibase 10.1126/science.1139831} {\bibfield
  {journal} {\bibinfo  {journal} {Science}\ }\textbf {\bibinfo {volume}
  {316}},\ \bibinfo {pages} {1312} (\bibinfo {year} {2007})}\BibitemShut
  {NoStop}%
\bibitem [{\citenamefont {Robledo}\ \emph {et~al.}(2011)\citenamefont
  {Robledo}, \citenamefont {Childress}, \citenamefont {Bernien}, \citenamefont
  {Hensen}, \citenamefont {Alkemade},\ and\ \citenamefont
  {Hanson}}]{robledo_high-fidelity_2011}%
  \BibitemOpen
  \bibfield  {author} {\bibinfo {author} {\bibfnamefont {L.}~\bibnamefont
  {Robledo}}, \bibinfo {author} {\bibfnamefont {L.}~\bibnamefont {Childress}},
  \bibinfo {author} {\bibfnamefont {H.}~\bibnamefont {Bernien}}, \bibinfo
  {author} {\bibfnamefont {B.}~\bibnamefont {Hensen}}, \bibinfo {author}
  {\bibfnamefont {P.~F.~A.}\ \bibnamefont {Alkemade}}, \ and\ \bibinfo {author}
  {\bibfnamefont {R.}~\bibnamefont {Hanson}},\ }\href {\doibase
  10.1038/nature10401} {\bibfield  {journal} {\bibinfo  {journal} {Nature}\
  }\textbf {\bibinfo {volume} {477}},\ \bibinfo {pages} {574} (\bibinfo {year}
  {2011})}\BibitemShut {NoStop}%
\bibitem [{\citenamefont {Liu}\ \emph {et~al.}(2013)\citenamefont {Liu},
  \citenamefont {Po}, \citenamefont {Du}, \citenamefont {Liu},\ and\
  \citenamefont {Pan}}]{liu_noise-resilient_2013}%
  \BibitemOpen
  \bibfield  {author} {\bibinfo {author} {\bibfnamefont {G.-Q.}\ \bibnamefont
  {Liu}}, \bibinfo {author} {\bibfnamefont {H.~C.}\ \bibnamefont {Po}},
  \bibinfo {author} {\bibfnamefont {J.}~\bibnamefont {Du}}, \bibinfo {author}
  {\bibfnamefont {R.-B.}\ \bibnamefont {Liu}}, \ and\ \bibinfo {author}
  {\bibfnamefont {X.-Y.}\ \bibnamefont {Pan}},\ }\href {\doibase
  10.1038/ncomms3254} {\bibfield  {journal} {\bibinfo  {journal} {Nat.
  Commun.}\ }\textbf {\bibinfo {volume} {4}},\ \bibinfo {pages} {2254}
  (\bibinfo {year} {2013})}\BibitemShut {NoStop}%
\bibitem [{\citenamefont {Waldherr}\ \emph {et~al.}(2014)\citenamefont
  {Waldherr}, \citenamefont {Wang}, \citenamefont {Zaiser}, \citenamefont
  {Jamali}, \citenamefont {Schulte-Herbr{\"u}ggen}, \citenamefont {Abe},
  \citenamefont {Ohshima}, \citenamefont {Isoya}, \citenamefont {Du},
  \citenamefont {Neumann},\ and\ \citenamefont
  {Wrachtrup}}]{waldherr_quantum_2014}%
  \BibitemOpen
  \bibfield  {author} {\bibinfo {author} {\bibfnamefont {G.}~\bibnamefont
  {Waldherr}}, \bibinfo {author} {\bibfnamefont {Y.}~\bibnamefont {Wang}},
  \bibinfo {author} {\bibfnamefont {S.}~\bibnamefont {Zaiser}}, \bibinfo
  {author} {\bibfnamefont {M.}~\bibnamefont {Jamali}}, \bibinfo {author}
  {\bibfnamefont {T.}~\bibnamefont {Schulte-Herbr{\"u}ggen}}, \bibinfo {author}
  {\bibfnamefont {H.}~\bibnamefont {Abe}}, \bibinfo {author} {\bibfnamefont
  {T.}~\bibnamefont {Ohshima}}, \bibinfo {author} {\bibfnamefont
  {J.}~\bibnamefont {Isoya}}, \bibinfo {author} {\bibfnamefont {J.~F.}\
  \bibnamefont {Du}}, \bibinfo {author} {\bibfnamefont {P.}~\bibnamefont
  {Neumann}}, \ and\ \bibinfo {author} {\bibfnamefont {J.}~\bibnamefont
  {Wrachtrup}},\ }\href {\doibase 10.1038/nature12919} {\bibfield  {journal}
  {\bibinfo  {journal} {Nature}\ }\textbf {\bibinfo {volume} {506}},\ \bibinfo
  {pages} {204} (\bibinfo {year} {2014})}\BibitemShut {NoStop}%
\bibitem [{\citenamefont {Maurer}\ \emph {et~al.}(2012)\citenamefont {Maurer},
  \citenamefont {Kucsko}, \citenamefont {Latta}, \citenamefont {Jiang},
  \citenamefont {Yao}, \citenamefont {Bennett}, \citenamefont {Pastawski},
  \citenamefont {Hunger}, \citenamefont {Chisholm}, \citenamefont {Markham},
  \citenamefont {Twitchen}, \citenamefont {Cirac},\ and\ \citenamefont
  {Lukin}}]{maurer_room-temperature_2012}%
  \BibitemOpen
  \bibfield  {author} {\bibinfo {author} {\bibfnamefont {P.~C.}\ \bibnamefont
  {Maurer}}, \bibinfo {author} {\bibfnamefont {G.}~\bibnamefont {Kucsko}},
  \bibinfo {author} {\bibfnamefont {C.}~\bibnamefont {Latta}}, \bibinfo
  {author} {\bibfnamefont {L.}~\bibnamefont {Jiang}}, \bibinfo {author}
  {\bibfnamefont {N.~Y.}\ \bibnamefont {Yao}}, \bibinfo {author} {\bibfnamefont
  {S.~D.}\ \bibnamefont {Bennett}}, \bibinfo {author} {\bibfnamefont
  {F.}~\bibnamefont {Pastawski}}, \bibinfo {author} {\bibfnamefont
  {D.}~\bibnamefont {Hunger}}, \bibinfo {author} {\bibfnamefont
  {N.}~\bibnamefont {Chisholm}}, \bibinfo {author} {\bibfnamefont
  {M.}~\bibnamefont {Markham}}, \bibinfo {author} {\bibfnamefont {D.~J.}\
  \bibnamefont {Twitchen}}, \bibinfo {author} {\bibfnamefont {J.~I.}\
  \bibnamefont {Cirac}}, \ and\ \bibinfo {author} {\bibfnamefont {M.~D.}\
  \bibnamefont {Lukin}},\ }\href {\doibase 10.1126/science.1220513} {\bibfield
  {journal} {\bibinfo  {journal} {Science}\ }\textbf {\bibinfo {volume}
  {336}},\ \bibinfo {pages} {1283} (\bibinfo {year} {2012})}\BibitemShut
  {NoStop}%
\bibitem [{\citenamefont {Barrett}\ and\ \citenamefont
  {Kok}(2005)}]{barrett_efficient_2005}%
  \BibitemOpen
  \bibfield  {author} {\bibinfo {author} {\bibfnamefont {S.~D.}\ \bibnamefont
  {Barrett}}\ and\ \bibinfo {author} {\bibfnamefont {P.}~\bibnamefont {Kok}},\
  }\href {\doibase 10.1103/PhysRevA.71.060310} {\bibfield  {journal} {\bibinfo
  {journal} {Phys. Rev. A}\ }\textbf {\bibinfo {volume} {71}},\ \bibinfo
  {pages} {060310} (\bibinfo {year} {2005})}\BibitemShut {NoStop}%
\bibitem [{\citenamefont {Taminiau}\ \emph {et~al.}(2014)\citenamefont
  {Taminiau}, \citenamefont {Cramer}, \citenamefont {van~der Sar},
  \citenamefont {Dobrovitski},\ and\ \citenamefont
  {Hanson}}]{taminiau_universal_2014}%
  \BibitemOpen
  \bibfield  {author} {\bibinfo {author} {\bibfnamefont {T.~H.}\ \bibnamefont
  {Taminiau}}, \bibinfo {author} {\bibfnamefont {J.}~\bibnamefont {Cramer}},
  \bibinfo {author} {\bibfnamefont {T.}~\bibnamefont {van~der Sar}}, \bibinfo
  {author} {\bibfnamefont {V.~V.}\ \bibnamefont {Dobrovitski}}, \ and\ \bibinfo
  {author} {\bibfnamefont {R.}~\bibnamefont {Hanson}},\ }\href {\doibase
  10.1038/nnano.2014.2} {\bibfield  {journal} {\bibinfo  {journal} {Nature
  Nanotech.}\ }\textbf {\bibinfo {volume} {9}},\ \bibinfo {pages} {171}
  (\bibinfo {year} {2014})}\BibitemShut {NoStop}%
\bibitem [{\citenamefont {Cramer}\ \emph {et~al.}(2015)\citenamefont {Cramer},
  \citenamefont {Kalb}, \citenamefont {Rol}, \citenamefont {Hensen},
  \citenamefont {Blok}, \citenamefont {Markham}, \citenamefont {Twitchen},
  \citenamefont {Hanson},\ and\ \citenamefont
  {Taminiau}}]{cramer_repeated_2015}%
  \BibitemOpen
  \bibfield  {author} {\bibinfo {author} {\bibfnamefont {J.}~\bibnamefont
  {Cramer}}, \bibinfo {author} {\bibfnamefont {N.}~\bibnamefont {Kalb}},
  \bibinfo {author} {\bibfnamefont {M.~A.}\ \bibnamefont {Rol}}, \bibinfo
  {author} {\bibfnamefont {B.}~\bibnamefont {Hensen}}, \bibinfo {author}
  {\bibfnamefont {M.~S.}\ \bibnamefont {Blok}}, \bibinfo {author}
  {\bibfnamefont {M.}~\bibnamefont {Markham}}, \bibinfo {author} {\bibfnamefont
  {D.~J.}\ \bibnamefont {Twitchen}}, \bibinfo {author} {\bibfnamefont
  {R.}~\bibnamefont {Hanson}}, \ and\ \bibinfo {author} {\bibfnamefont {T.~H.}\
  \bibnamefont {Taminiau}},\ }\href {http://arxiv.org/abs/1508.01388}
  {\bibfield  {journal} {\bibinfo  {journal} {arXiv:1508.01388}\ } (\bibinfo
  {year} {2015})}\BibitemShut {NoStop}%
\bibitem [{\citenamefont {Taminiau}\ \emph {et~al.}(2012)\citenamefont
  {Taminiau}, \citenamefont {Wagenaar}, \citenamefont {van~der Sar},
  \citenamefont {Jelezko}, \citenamefont {Dobrovitski},\ and\ \citenamefont
  {Hanson}}]{taminiau_detection_2012}%
  \BibitemOpen
  \bibfield  {author} {\bibinfo {author} {\bibfnamefont {T.~H.}\ \bibnamefont
  {Taminiau}}, \bibinfo {author} {\bibfnamefont {J.~J.~T.}\ \bibnamefont
  {Wagenaar}}, \bibinfo {author} {\bibfnamefont {T.}~\bibnamefont {van~der
  Sar}}, \bibinfo {author} {\bibfnamefont {F.}~\bibnamefont {Jelezko}},
  \bibinfo {author} {\bibfnamefont {V.~V.}\ \bibnamefont {Dobrovitski}}, \ and\
  \bibinfo {author} {\bibfnamefont {R.}~\bibnamefont {Hanson}},\ }\href
  {\doibase 10.1103/PhysRevLett.109.137602} {\bibfield  {journal} {\bibinfo
  {journal} {Phys. Rev. Lett.}\ }\textbf {\bibinfo {volume} {109}},\ \bibinfo
  {pages} {137602} (\bibinfo {year} {2012})}\BibitemShut {NoStop}%
\bibitem [{\citenamefont {Jiang}\ \emph {et~al.}(2008)\citenamefont {Jiang},
  \citenamefont {Dutt}, \citenamefont {Togan}, \citenamefont {Childress},
  \citenamefont {Cappellaro}, \citenamefont {Taylor},\ and\ \citenamefont
  {Lukin}}]{jiang_coherence_2008}%
  \BibitemOpen
  \bibfield  {author} {\bibinfo {author} {\bibfnamefont {L.}~\bibnamefont
  {Jiang}}, \bibinfo {author} {\bibfnamefont {M.~V.~G.}\ \bibnamefont {Dutt}},
  \bibinfo {author} {\bibfnamefont {E.}~\bibnamefont {Togan}}, \bibinfo
  {author} {\bibfnamefont {L.}~\bibnamefont {Childress}}, \bibinfo {author}
  {\bibfnamefont {P.}~\bibnamefont {Cappellaro}}, \bibinfo {author}
  {\bibfnamefont {J.~M.}\ \bibnamefont {Taylor}}, \ and\ \bibinfo {author}
  {\bibfnamefont {M.~D.}\ \bibnamefont {Lukin}},\ }\href {\doibase
  10.1103/PhysRevLett.100.073001} {\bibfield  {journal} {\bibinfo  {journal}
  {Phys. Rev. Lett.}\ }\textbf {\bibinfo {volume} {100}},\ \bibinfo {pages}
  {073001} (\bibinfo {year} {2008})}\BibitemShut {NoStop}%
\bibitem [{\citenamefont {Blok}\ \emph {et~al.}(2015)\citenamefont {Blok},
  \citenamefont {Kalb}, \citenamefont {Reiserer}, \citenamefont {Taminiau},\
  and\ \citenamefont {Hanson}}]{blok_towards_2015}%
  \BibitemOpen
  \bibfield  {author} {\bibinfo {author} {\bibfnamefont {M.~S.}\ \bibnamefont
  {Blok}}, \bibinfo {author} {\bibfnamefont {N.}~\bibnamefont {Kalb}}, \bibinfo
  {author} {\bibfnamefont {A.}~\bibnamefont {Reiserer}}, \bibinfo {author}
  {\bibfnamefont {T.~H.}\ \bibnamefont {Taminiau}}, \ and\ \bibinfo {author}
  {\bibfnamefont {R.}~\bibnamefont {Hanson}},\ }\href {\doibase
  10.1039/C5FD00113G} {\bibfield  {journal} {\bibinfo  {journal} {Faraday
  Discuss.}\ }\textbf {\bibinfo {volume} {184}},\ \bibinfo {pages} {173}
  (\bibinfo {year} {2015})}\BibitemShut {NoStop}%
\bibitem [{\citenamefont {Goldman}\ \emph {et~al.}(2015)\citenamefont
  {Goldman}, \citenamefont {Sipahigil}, \citenamefont {Doherty}, \citenamefont
  {Yao}, \citenamefont {Bennett}, \citenamefont {Markham}, \citenamefont
  {Twitchen}, \citenamefont {Manson}, \citenamefont {Kubanek},\ and\
  \citenamefont {Lukin}}]{goldman_phonon-induced_2015}%
  \BibitemOpen
  \bibfield  {author} {\bibinfo {author} {\bibfnamefont {M.~L.}\ \bibnamefont
  {Goldman}}, \bibinfo {author} {\bibfnamefont {A.}~\bibnamefont {Sipahigil}},
  \bibinfo {author} {\bibfnamefont {M.~W.}\ \bibnamefont {Doherty}}, \bibinfo
  {author} {\bibfnamefont {N.~Y.}\ \bibnamefont {Yao}}, \bibinfo {author}
  {\bibfnamefont {S.~D.}\ \bibnamefont {Bennett}}, \bibinfo {author}
  {\bibfnamefont {M.}~\bibnamefont {Markham}}, \bibinfo {author} {\bibfnamefont
  {D.~J.}\ \bibnamefont {Twitchen}}, \bibinfo {author} {\bibfnamefont {N.~B.}\
  \bibnamefont {Manson}}, \bibinfo {author} {\bibfnamefont {A.}~\bibnamefont
  {Kubanek}}, \ and\ \bibinfo {author} {\bibfnamefont {M.~D.}\ \bibnamefont
  {Lukin}},\ }\href {\doibase 10.1103/PhysRevLett.114.145502} {\bibfield
  {journal} {\bibinfo  {journal} {Phys. Rev. Lett.}\ }\textbf {\bibinfo
  {volume} {114}},\ \bibinfo {pages} {145502} (\bibinfo {year}
  {2015})}\BibitemShut {NoStop}%
\bibitem [{\citenamefont {Doherty}\ \emph {et~al.}(2013)\citenamefont
  {Doherty}, \citenamefont {Manson}, \citenamefont {Delaney}, \citenamefont
  {Jelezko}, \citenamefont {Wrachtrup},\ and\ \citenamefont
  {Hollenberg}}]{doherty_nitrogen-vacancy_2013}%
  \BibitemOpen
  \bibfield  {author} {\bibinfo {author} {\bibfnamefont {M.~W.}\ \bibnamefont
  {Doherty}}, \bibinfo {author} {\bibfnamefont {N.~B.}\ \bibnamefont {Manson}},
  \bibinfo {author} {\bibfnamefont {P.}~\bibnamefont {Delaney}}, \bibinfo
  {author} {\bibfnamefont {F.}~\bibnamefont {Jelezko}}, \bibinfo {author}
  {\bibfnamefont {J.}~\bibnamefont {Wrachtrup}}, \ and\ \bibinfo {author}
  {\bibfnamefont {L.~C.~L.}\ \bibnamefont {Hollenberg}},\ }\href {\doibase
  10.1016/j.physrep.2013.02.001} {\bibfield  {journal} {\bibinfo  {journal}
  {Physics Reports}\ }\textbf {\bibinfo {volume} {528}},\ \bibinfo {pages} {1}
  (\bibinfo {year} {2013})}\BibitemShut {NoStop}%
\bibitem [{\citenamefont {Robledo}\ \emph {et~al.}(2010)\citenamefont
  {Robledo}, \citenamefont {Bernien}, \citenamefont {van Weperen},\ and\
  \citenamefont {Hanson}}]{robledo_control_2010}%
  \BibitemOpen
  \bibfield  {author} {\bibinfo {author} {\bibfnamefont {L.}~\bibnamefont
  {Robledo}}, \bibinfo {author} {\bibfnamefont {H.}~\bibnamefont {Bernien}},
  \bibinfo {author} {\bibfnamefont {I.}~\bibnamefont {van Weperen}}, \ and\
  \bibinfo {author} {\bibfnamefont {R.}~\bibnamefont {Hanson}},\ }\href
  {\doibase 10.1103/PhysRevLett.105.177403} {\bibfield  {journal} {\bibinfo
  {journal} {Phys. Rev. Lett.}\ }\textbf {\bibinfo {volume} {105}},\ \bibinfo
  {pages} {177403} (\bibinfo {year} {2010})}\BibitemShut {NoStop}%
\bibitem [{\citenamefont {Vandersypen}\ and\ \citenamefont
  {Chuang}(2005)}]{vandersypen_nmr_2005}%
  \BibitemOpen
  \bibfield  {author} {\bibinfo {author} {\bibfnamefont {L.~M.~K.}\
  \bibnamefont {Vandersypen}}\ and\ \bibinfo {author} {\bibfnamefont {I.~L.}\
  \bibnamefont {Chuang}},\ }\href {\doibase 10.1103/RevModPhys.76.1037}
  {\bibfield  {journal} {\bibinfo  {journal} {Rev. Mod. Phys.}\ }\textbf
  {\bibinfo {volume} {76}},\ \bibinfo {pages} {1037} (\bibinfo {year}
  {2005})}\BibitemShut {NoStop}%
\bibitem [{\citenamefont {Lidar}\ \emph {et~al.}(1998)\citenamefont {Lidar},
  \citenamefont {Chuang},\ and\ \citenamefont
  {Whaley}}]{lidar_decoherence-free_1998}%
  \BibitemOpen
  \bibfield  {author} {\bibinfo {author} {\bibfnamefont {D.~A.}\ \bibnamefont
  {Lidar}}, \bibinfo {author} {\bibfnamefont {I.~L.}\ \bibnamefont {Chuang}}, \
  and\ \bibinfo {author} {\bibfnamefont {K.~B.}\ \bibnamefont {Whaley}},\
  }\href {\doibase 10.1103/PhysRevLett.81.2594} {\bibfield  {journal} {\bibinfo
   {journal} {Phys. Rev. Lett.}\ }\textbf {\bibinfo {volume} {81}},\ \bibinfo
  {pages} {2594} (\bibinfo {year} {1998})}\BibitemShut {NoStop}%
\bibitem [{\citenamefont {Awschalom}\ \emph {et~al.}(2013)\citenamefont
  {Awschalom}, \citenamefont {Bassett}, \citenamefont {Dzurak}, \citenamefont
  {Hu},\ and\ \citenamefont {Petta}}]{awschalom_quantum_2013}%
  \BibitemOpen
  \bibfield  {author} {\bibinfo {author} {\bibfnamefont {D.~D.}\ \bibnamefont
  {Awschalom}}, \bibinfo {author} {\bibfnamefont {L.~C.}\ \bibnamefont
  {Bassett}}, \bibinfo {author} {\bibfnamefont {A.~S.}\ \bibnamefont {Dzurak}},
  \bibinfo {author} {\bibfnamefont {E.~L.}\ \bibnamefont {Hu}}, \ and\ \bibinfo
  {author} {\bibfnamefont {J.~R.}\ \bibnamefont {Petta}},\ }\href {\doibase
  10.1126/science.1231364} {\bibfield  {journal} {\bibinfo  {journal}
  {Science}\ }\textbf {\bibinfo {volume} {339}},\ \bibinfo {pages} {1174}
  (\bibinfo {year} {2013})}\BibitemShut {NoStop}%
\bibitem [{\citenamefont {Purcell}(1946)}]{purcell_spontaneous_1946}%
  \BibitemOpen
  \bibfield  {author} {\bibinfo {author} {\bibfnamefont {E.~M.}\ \bibnamefont
  {Purcell}},\ }\href {http://prola.aps.org/pdf/PR/v69/i11-12/p674_2}
  {\bibfield  {journal} {\bibinfo  {journal} {Phys. Rev.}\ }\textbf {\bibinfo
  {volume} {69}},\ \bibinfo {pages} {681} (\bibinfo {year} {1946})}\BibitemShut
  {NoStop}%
\bibitem [{\citenamefont {Kalb}\ \emph {et~al.}(2015)\citenamefont {Kalb},
  \citenamefont {Reiserer}, \citenamefont {Ritter},\ and\ \citenamefont
  {Rempe}}]{kalb_heralded_2015}%
  \BibitemOpen
  \bibfield  {author} {\bibinfo {author} {\bibfnamefont {N.}~\bibnamefont
  {Kalb}}, \bibinfo {author} {\bibfnamefont {A.}~\bibnamefont {Reiserer}},
  \bibinfo {author} {\bibfnamefont {S.}~\bibnamefont {Ritter}}, \ and\ \bibinfo
  {author} {\bibfnamefont {G.}~\bibnamefont {Rempe}},\ }\href {\doibase
  10.1103/PhysRevLett.114.220501} {\bibfield  {journal} {\bibinfo  {journal}
  {Phys. Rev. Lett.}\ }\textbf {\bibinfo {volume} {114}},\ \bibinfo {pages}
  {220501} (\bibinfo {year} {2015})}\BibitemShut {NoStop}%
\bibitem [{\citenamefont {Yang}\ \emph {et~al.}(2015)\citenamefont {Yang},
  \citenamefont {Wang}, \citenamefont {Rao}, \citenamefont {Tran},
  \citenamefont {Momenzadeh}, \citenamefont {Nagy}, \citenamefont {Markham},
  \citenamefont {Twitchen}, \citenamefont {Wang}, \citenamefont {Yang},
  \citenamefont {Stoehr}, \citenamefont {Neumann}, \citenamefont {Kosaka},\
  and\ \citenamefont {Wrachtrup}}]{yang_high_2015}%
  \BibitemOpen
  \bibfield  {author} {\bibinfo {author} {\bibfnamefont {S.}~\bibnamefont
  {Yang}}, \bibinfo {author} {\bibfnamefont {Y.}~\bibnamefont {Wang}}, \bibinfo
  {author} {\bibfnamefont {D.~D.~B.}\ \bibnamefont {Rao}}, \bibinfo {author}
  {\bibfnamefont {T.~H.}\ \bibnamefont {Tran}}, \bibinfo {author}
  {\bibfnamefont {S.~A.}\ \bibnamefont {Momenzadeh}}, \bibinfo {author}
  {\bibfnamefont {R.}~\bibnamefont {Nagy}}, \bibinfo {author} {\bibfnamefont
  {M.}~\bibnamefont {Markham}}, \bibinfo {author} {\bibfnamefont {D.~J.}\
  \bibnamefont {Twitchen}}, \bibinfo {author} {\bibfnamefont {P.}~\bibnamefont
  {Wang}}, \bibinfo {author} {\bibfnamefont {W.}~\bibnamefont {Yang}}, \bibinfo
  {author} {\bibfnamefont {R.}~\bibnamefont {Stoehr}}, \bibinfo {author}
  {\bibfnamefont {P.}~\bibnamefont {Neumann}}, \bibinfo {author} {\bibfnamefont
  {H.}~\bibnamefont {Kosaka}}, \ and\ \bibinfo {author} {\bibfnamefont
  {J.}~\bibnamefont {Wrachtrup}},\ }\href {http://arxiv.org/abs/1511.04939}
  {\bibfield  {journal} {\bibinfo  {journal} {arXiv:1511.04939}\ } (\bibinfo
  {year} {2015})}\BibitemShut {NoStop}%
\bibitem [{\citenamefont {Jones}\ \emph {et~al.}(2015)\citenamefont {Jones},
  \citenamefont {Kim}, \citenamefont {Rakher}, \citenamefont {Kwiat},\ and\
  \citenamefont {Ladd}}]{jones_design_2015}%
  \BibitemOpen
  \bibfield  {author} {\bibinfo {author} {\bibfnamefont {C.}~\bibnamefont
  {Jones}}, \bibinfo {author} {\bibfnamefont {D.}~\bibnamefont {Kim}}, \bibinfo
  {author} {\bibfnamefont {M.~T.}\ \bibnamefont {Rakher}}, \bibinfo {author}
  {\bibfnamefont {P.~G.}\ \bibnamefont {Kwiat}}, \ and\ \bibinfo {author}
  {\bibfnamefont {T.~D.}\ \bibnamefont {Ladd}},\ }\href
  {http://arxiv.org/abs/1505.01536} {\bibfield  {journal} {\bibinfo  {journal}
  {arXiv:1505.01536}\ } (\bibinfo {year} {2015})}\BibitemShut {NoStop}%
\bibitem [{\citenamefont {Campbell}\ and\ \citenamefont
  {Benjamin}(2008)}]{campbell_measurement-based_2008}%
  \BibitemOpen
  \bibfield  {author} {\bibinfo {author} {\bibfnamefont {E.~T.}\ \bibnamefont
  {Campbell}}\ and\ \bibinfo {author} {\bibfnamefont {S.~C.}\ \bibnamefont
  {Benjamin}},\ }\href {\doibase 10.1103/PhysRevLett.101.130502} {\bibfield
  {journal} {\bibinfo  {journal} {Phys. Rev. Lett.}\ }\textbf {\bibinfo
  {volume} {101}},\ \bibinfo {pages} {130502} (\bibinfo {year}
  {2008})}\BibitemShut {NoStop}%
\end{thebibliography}%

\end{document}